\documentclass[RNAAS,twocolumn]{aastex62}
\usepackage{graphicx,url,amssymb,amsmath,color,units,wasysym,epsfig,epstopdf,enumerate}

\usepackage{newtxtext,newtxmath}
\usepackage[T1]{fontenc}
\usepackage{ae,aecompl}
\usepackage[left]{lineno}
\usepackage[normalem]{ulem}
\usepackage[bottom]{footmisc}

\newcommand{\Nevents}{$62$}
\newcommand{\Necc}{four}

\renewcommand{\sc}[1]{\textsc{#1}}

\newcommand{\new}[1]{{#1}}
\newcommand{\newnew}[1]{{#1}}

\newcommand{\SPA}{School of Physics and Astronomy, Monash University, Clayton VIC 3800, Australia}
\newcommand{\OzGravMonash}{OzGrav: The ARC Centre of Excellence for Gravitational Wave Discovery, Clayton VIC 3800, Australia}
\newcommand{\DAMTP}{Department of Applied Mathematics and Theoretical Physics, Cambridge CB3 0WA, United Kingdom}

\begin{document}

\title{Four eccentric mergers increase the evidence that LIGO--Virgo--KAGRA's binary black holes form dynamically}

\author{Isobel Romero-Shaw}
    \email{ir346@cam.ac.uk}
\affiliation{\SPA}
\affiliation{\OzGravMonash}
\affiliation{\DAMTP}

\author{Paul D. Lasky}
\affiliation{\SPA}
\affiliation{\OzGravMonash}

\author{Eric Thrane}
\affiliation{\SPA}

\begin{abstract}
The growing population of compact binary mergers detected with gravitational waves contains multiple events that are challenging to explain through isolated binary evolution.
Such events have higher masses than are expected in isolated binaries, component spin-tilt angles that are misaligned, and/or non-negligible orbital eccentricities.
We investigate the orbital eccentricities of \Nevents~binary black hole candidates from the third gravitational-wave transient catalogue of the LIGO-Virgo-KAGRA Collaboration \new{with an aligned-spin, moderate-eccentricity waveform model.} 
\new{Within this framework, we find} that at least four of these events show significant support for eccentricity $e_{10} \geq 0.1$ at a gravitational-wave frequency of $10$~Hz \new{($> 60\%$ credibility, under a log-uniform eccentricity prior that spans the range $10^{-4} < e_{10} < 0.2$)}.
Two of these events are new additions to the population: GW191109 and GW200208\_22.
If the four eccentric candidates are truly eccentric, our results suggest that densely-populated star clusters may produce 100\% of the observed mergers.
However, it remains likely that other formation environments with higher yields of eccentric mergers---for example, active galactic nuclei---also contribute.
We estimate that we will be able to confidently distinguish which formation channel dominates the eccentric merger rate after $\gtrsim 80$ detections of events with $e_{10} \geq 0.05$ at LIGO--Virgo sensitivity, with only $\sim 5$ detectably-eccentric events required to distinguish formation channels with third-generation gravitational-wave detectors.
\end{abstract}


\section{Introduction}
\label{sec:intro}
The LIGO-Virgo-KAGRA (LVK) collaboration has so far reported 90 gravitational-wave signals of probable ($\geq 50\%$ credible) astrophysical origin \citep{GWTC-1, GWTC-2, GWTC-2-1, GWTC-3}, all consistent with coming the mergers of a compact binary: a binary black hole (BBH), binary neutron star (BNS) or neutron star--black hole binary (NSBH). 
The provenance of these compact-object binary mergers is an open question in gravitational-wave astrophysics.
In order for an isolated pair of stars to merge as a compact binary on an observable timescale, it must undergo specific evolutionary scenarios.
Typically, isolated binaries must harden either through Roche-lobe overflow mass transfer \citep{vandeHuevel:2017:stableMT, Neijssel:2019:stableMT, Bavera2020, Olejak:2021:stableMT, Gallegos-Garcia:2021:stableMT} or common-envelope evolution \citep{Livio88, Bethe98, Ivanova13, Kruckow16}, or be born with a small enough separation that chemically-homogeneous evolution is possible \citep[e.g.,][]{deMink10, deMink16, Marchant16}.
Compact objects can instead be driven to merge rapidly by dynamical interactions, which can happen in populous environments like star clusters \citep[e.g.,][]{Rodriguez18a, Rodriguez18b, Samsing18, Fragione:2020:GW190521StarClusersHierarcical}.

The different processes that facilitate the merger of a binary leave their signature on the resulting gravitational-wave signal.
Multiple studies have shown how the compact-object masses, spins, and orbital eccentricities inferred from the signal may act as identifiers of different formation scenarios, both for individual events and for the contribution of different formation pathways to the entire population~\citep[e.g.,][]{Vitale15, Stevenson15bqa, Gerosa17, Bavera2020, Mapelli:2020:review, Sedda:2020:fingerprints, Zevin:2020:channels, Fragione:2022:NSCs}.
In addition, the redshift evolution of the merger rate should contain distinct contributions from different formation channels \citep[e.g.,][]{Mandic:2016:stochastic-background-primordial, Francolini2021}, although this will only be resolvable after $\mathcal{O}(100)$ detections \citep{Fishbach:2018:mergerrate}.

Non-zero orbital eccentricity is arguably the most robust signature of dynamical formation.\footnote{Lidov-Kozai resonant oscillations can drive up the eccentricity of a merging binary in an isolated triple system \citep{Lidov62, Kozai62, Naoz2016, Antonini17}, but the relative contribution of this channel to the observable eccentric merger rate is thought to be low even if optimistically low black-hole natal kicks and metallicities are assumed \citep{Silsbee16, RodriguezAntonini2018}.}
A binary undergoing isolated evolution is expected to circularise before its gravitational-wave frequency reaches the start of the LVK frequency band at $10$~Hz \citep{Peters64}.
In contrast, dynamically-induced mergers often merge so rapidly that they retain non-negligible orbital eccentricity at $10$~Hz, a quantity that we refer to as $e_{10}$.
Robust predictions exist for the distribution of binary eccentricities in dense star clusters \citep[e.g.,][]{Samsing17, SamsingRamirez17, SamsingDOrazio18, Zevin18, Rodriguez18a, Rodriguez18b}, with $\sim 5\%$ of mergers in these environments retaining $e_{10} \geq 0.1$.
Expectations from galactic nuclei and active galactic nuclei (AGN) disks are also becoming clearer, with up to $70\%$ of mergers in AGN disks thought to retain $e_{10} \geq 0.1$ \citep[e.g.][]{Samsing:2020:AGNeccentric, Tagawa:2021:AGN, GondanKocsis2021, Vajpeyi:2021:AGN}.

\new{Existing detectors are sensitive to eccentricities $e_{10} \gtrsim 0.05$ for BBH mergers \citep{Lower18, Romero-Shaw:2019:GWTC-1-ecc}. 
Signal detection currently depends on achieving a high signal-to-noise ratio when the data are match-filtered using a quasi-circular signal template, so eccentric signals have reduced detectability compared to quasi-circular signals. 
Roughly half of the $\sim 7\%$ of mergers in dense star clusters with $e_{10} \geq 0.05$ are recoverable with such a search \citep{Zevin:2021:seleccentricity}.
Therefore, accounting for current detector sensitivities and the loss of signal-to-noise power when using quasi-circular templates to search for eccentric signals, we expect to be able to measure the eccentricities of $\sim 4\%$ of mergers from dense star clusters.
This percentage is an underestimate for the true fraction of detectably-eccentric sources recovered, since it is based on the overlap between eccentric and quasi-circular waveforms with otherwise identical parameters.}

Although non-zero orbital eccentricity is one calling card of dynamical formation, the component masses of a compact-object merger may also help to distinguish its origins. 
While pair-instability supernovae prevent black holes forming between around $\sim60$ and $\sim130$~M$_\odot$ in isolation \citep{HegerWoosley02, Belczynski16, Marchant16, Woosley17, Fishbach17}, dynamical environments can build massive black holes through hierarchical mergers or accretion \citep[e.g.,][]{Fishbach17a, Rodriguez19, Gerosa:2019:EscapeSpeed, Anagnostou:2020:GW190521, Kimball:2020:hierarchical, Kremer2020, Samsing:2020:massgap, Fragione:2020:RepeatedMargersAndEjections, Gerosa:2021:Hierachical, Banerjee:2021:YoungOpenClusters, Zevin:2022:ClusterCatastrophe}.
Nonetheless, the uncertain limits and range of the pair-instability mass gap~\citep{Sakstein:2020:0521, Belczynski:2020:0521, Farmer2019, Woosley2021, Ziegler2021} reduce the efficacy of compact-object mass as an identifier of formation channel.

The spin \textit{directions} of a binary's components can indicate its formation mechanism \citep{Farr17, Stevenson17dlk, TalbotThrane17}.
Isolated binaries are expected to have spins approximately aligned with the binary angular momentum vector \citep[e.g.,][]{Campanelli06, OShaughnessy17, Gerosa:2018wbw, Kalogera2000}, while dynamically-assembled pairs in spherically-symmetric environments should have an isotropic distribution of relative spin tilts \citep[e.g.,][]{Rodriguez16}.
However, mergers with aligned spins are not necessarily of isolated origin: gas torques in AGN disks can align component spins \citep{Bogdanociv2007} if the timescale for dynamical interaction is sufficiently long \citep{LiuLai17, Tagawa:2020:AGNdiskspin}, and dynamically-assembled binaries in open clusters can have aligned spins because few dynamical encounters can occur before merger \citep{Trani:2021:open-cluster-spin-tilt}.

Black hole spin \textit{magnitudes}, too, can distinguish binary formation mechanisms.
Merger products should develop high spins as they accumulate angular momentum through repeated mergers, so high spin magnitude can indicate dynamical formation \citep{Fishbach17a, Kimball:2020:hierarchical, Tagawa:2021:hierarchicalspin}, since the spins of black holes that form via stellar collapse are typically expected to be small. 
However, chemically homogeneous evolution  \citep{Marchant16, MandelDeMink2016, Qin:2019:HMXB-spins}, mass transfer and/or tidal locking in tight binaries \citep{Valsecchi10, Qin:2019:HMXB-spins, Neijssel:2021:CygX1, Izzard2003FormationRO, Belczynski2020, Bavera2020, Zevin:2022:SuspiciousSiblings, Broekgaarden:2022:MassRatioReversal}, and differential rotation between the stellar core and envelope \citep{Hirschi2005} can also produce rapidly-spinning black holes.

In addition to those proceeding via isolated evolution or dynamical formation, mergers can also occur between primordial black holes that form through direct collapse of density fluctuations in the early Universe.
While primordial black holes can have masses and spins that mimic those of isolated or dynamical black holes, with no upper or lower mass limit and preferentially zero spin (unless spun-up by accretion over the majority of cosmic history), primordial mergers should have zero eccentricity at detection \citep{Green2021, Francolini2021}.
Non-zero orbital eccentricity is also, therefore, a reliable way to rule out the primordial binary hypothesis.

Within the growing population of LVK observations are events that are challenging to explain through isolated stellar evolution.
The first event to breach the pair-instability mass gap at $\geq 90\%$ confidence was GW190521 \citep{GW190521}, which also exhibited signs of spin-induced precession and/or orbital eccentricity greater than $0.1$ at 10~Hz \citep{Romero-Shaw:2020:GW190521, Gayathri, JuanBosonStars, Gamba:2021:GW190521}.
A second binary black hole merger, GW190620, also supports a dynamical formation hypothesis, with $e_{10} \geq 0.05$ at $74\%$ credibility \citep{Romero-Shaw:2021:GWTC-2-ecc}.
New events that support the dynamical formation hypothesis have emerged in GWTC-2.1 and GWTC-3.
These include additional upper mass-gap events, such as GW190426\_19 \citep{GWTC-2-1} and GW200220\_06 \citep{GWTC-3}, and events consistent with having negatively-aligned or substantially misaligned component spins, such as GW191109 and GW200129 \citep{GWTC-3, GW200129}.\footnote{The strength of the support for anti-aligned or misaligned spins in GW200129 and GW191109 is contested by \citet{Payne:2022:200129} and Tong et al. in prep., respectively, who show that this support may be highly dependent on the data cleaning methods used.}
On top of these individual-event clues, population-level hints of dynamical formation come from evidence for hierarchical mergers \citep{Kimball:2021} and the mass-gap-encroaching shape of the inferred mass distribution. 
Hints of misaligned spins (negative effective spin parameter $\chi_{\rm eff}$) have been claimed at the population level \citep{GWTC-3-RnP}; some follow-up studies show that the population is consistent with the majority of binaries having non-zero and misaligned component spins, while others argue that the population is consistent with a majority of binaries having $\chi_{\rm eff} = 0$ and only a small subset having significant positive $\chi_{\rm eff}$ \citep{Roulet:2021:spin, Galaudage:2021:spin}.

Eccentricity is not included in the gravitational-waveform models used by the LVK to produce the inferences reported in their catalogues, because incorporating the effects of eccentricity makes physically-accurate waveform models too slow for conventional inference methods.
In \citet{Romero-Shaw:2019:GWTC-1-ecc, Romero-Shaw:2020:GW190425, Romero-Shaw:2020:GW190521, Romero-Shaw:2021:GWTC-2-ecc}, we used an efficient reweighting method to obtain measurements of the orbital eccentricity of gravitational-wave sources up to and including the second LVK gravitational-wave transient catalogue, GWTC-2.
In this work, we use the same method to analyse additional binary black hole candidates from the most recent \new{updates to the catalogue of LVK events: GWTC-2.1 \citep{GWTC-2-1} and GWTC-3 \citep{GWTC-3}}.
\new{Our results come with important caveats: we cannot distinguish eccentricity from spin-induced precession, our analysis does not include higher-order modes, and we are limited to studying moderate eccentricities ($e_{10} \leq 0.2$) and restricted spins ($\chi_{\rm eff} \leq 0.6$).}
\new{The limitations of our method are explained in detail in Section \ref{sec:methods}.}
\new{Within our analysis framework,} we report an additional two binaries with significant support for $e_{10} \geq 0.05$ \new{($> 60\%$ credibility)}, adding to the building \new{circumstantial} evidence for a dynamically-formed subset within the observed mergers.

This paper is structured as follows.
In Section \ref{sec:methods} we describe our methodology, and note its limitations.
In Section \ref{sec:results} we present results from our analysis of new events from GWTC-2.1 and GWTC-3, taking the total number of BBH candidates investigated for signatures of eccentricity to \Nevents.
Two events, GW191109 and GW200208\_22, have $\geq 70\%$ of their posterior support at $e_{10} \geq 0.05$ and have inconclusive but positive ln Bayes factors in favour of the eccentric hypothesis, with ln~$\mathcal{B} (e_{10} \geq 0.05)\gtrsim1.4$\new{, where we use the convention that ln~$\mathcal{B} \geq 8$ constitutes ``strong'' evidence.}
We present analyses of eccentricity at the population level in Section \ref{sec:population}, and demonstrate that $\gtrsim 80$ detectably-eccentric mergers are required to confidently distinguish different dynamical formation scenarios at current detector sensitivity.
In Section \ref{sec:conclusion}, we conclude with some final thoughts.
Results for events with negligible eccentricity are provided in Appendix \ref{sec:noneccentric}.

\section{Method}\label{sec:methods}
We use a reweighting method \citep[see][]{Payne, Romero-Shaw:2019:GWTC-1-ecc} to efficiently calculate posterior probability distributions using the aligned-spin eccentric waveform model \texttt{SEOBNRE} \citep{SEOBNRE}.
First, we run an importance-sampling step, performing Bayesian inference using \texttt{bilby} and the \texttt{bilby\_pipe} pipeline \citep{bilby, Romero-Shaw:2020:Bilby}.
We run five parallel analyses with unique seeds for each event with the \texttt{dynesty} sampler \citep{dynesty}, utilising spin-aligned quasi-circular model \texttt{IMRPhenomD} \citep{Khan15} as the `proposal' model.
For these initial analyses, we use $1000$ live points, $100$ walks and $10$ auto-correlation times.
For follow-up analyses on eccentric candidates, we use $4000$ live points and $200$ walks.
We use a sampling rate of $4096$~Hz and a reference frequency of $10$~Hz for every event.
We analyse publicly-available data from GWTC-2.1 and GWTC-3 \citep{GWTC-2-1, GWTC-3, GWOSC:GWTC-2-1, GWOSC:GWTC-3} and use detector noise curves generated using BayesWave \citep{Cornish:2014kda, Littenberg:2014oda}.

We reweight the proposal samples obtained in the initial step to our `target' model: \texttt{SEOBNRE}, a spin-aligned eccentric waveform approximant containing the inspiral, merger and ringdown sections of the signal.
Since this is a time-domain model, we use a Fourier transform to obtain frequency-domain waveforms for use in the likelihood, softening the abrupt start of the time-domain inspiral using a half-Tukey window to avoid spectral leakage.

We use standard priors on right ascension $\alpha$, declination $\delta$, source inclination $\theta_{\rm JN}$, polarisation $\phi$, coalescence phase $\psi$ and geocent time $t_0$.
Our prior on mass ratio $q$ is uniform between $0.125$ and $1$, and our priors on the aligned component spin magnitudes $\chi_1$, $\chi_2$ are capped at the \texttt{SEOBNRE} maximum of $\pm 0.6$.
We use uniform priors on chirp mass $\mathcal{M}$ and priors on luminosity distance $d_{\rm L}$ that are uniform in the source frame.
\new{When reconstructing the eccentricity distribution with \texttt{SEOBNRE}, we employ a log-uniform prior on eccentricity, which covers the range $10^{-4} \leq e_{10} \leq 0.2$.}
We marginalise over phase and coalescence time to mitigate definitional differences between the two models.

Including all events in GWTC-3, we report results for \Nevents~BBH candidates in total. \new{In this work, we present analyses of events added to the catalogue in GWTC-2.1 and GWTC-3; for analyses of events added to the catalogue in earlier LVK publications using the same method as used here, see \citet{Romero-Shaw:2019:GWTC-1-ecc} and \citet{Romero-Shaw:2021:GWTC-2-ecc}.}  We reserve detailed analyses of other events, including binaries that contain neutron stars, for future work.\footnote{BNS mergers have so far been found to be consistent with quasi-circularity \citep{Romero-Shaw:2020:GW190425, Lenon:2020:GW190425}}. The events that we do \textit{not} discuss in this work are:

\begin{itemize}
    \item \textbf{Events for which the reweighting process failed to achieve an adequate effective sample size.} The number of effective samples in a posterior distribution after reweighting is 
    \begin{equation}
        \label{eq:neff}
        n_{\rm eff} = \frac{\left(\sum_{i=1}^{n} w_{i}\right)^2}{\sum_{i=1}^{n} w_{i}^{2}}.
    \end{equation}
    We deem any events with $n_{\rm eff} < 100$ undersampled, and do not include them in this work. \new{Potential causes of undersampling are explored in Appendix 
    \ref{sec:undersampled}}.
    
    \item \textbf{Binaries likely to contain at least one neutron star.} These include those confidently designated as binary neutron stars (BNS; GW170817, GW190425), neutron star--black hole (NSBH) binaries (GW191219, GW200105, GW200115), and events with secondaries of ambiguous mass that may be black holes or neutron stars (GW190814, GW190917, GW200210). We neglect BNS and likely NSBH events from the analysis presented in the main body of this paper because their formation mechanisms may be drastically different from those of BBH mergers, and we wish to make statements about the formation channels that produce the BBH mergers in our population. When studied with the reweighting method, we find that GW190814A fails to achieve a high reweighting efficiency; this is expected, since higher-order modes are known to be present in this signal, and both models used in this work incorporate only the $(2, 2)$ mode \citep{GW190814}. GW190412, which is thought to be a unequal-mass black hole binary with mass ratio $q \sim 0.27$, also has higher-order mode content, and also ends up undersampled. 
    \new{We do not provide posterior probability distributions for GW190917 and GW200210, despite their adequate sample size after reweighting; their mass ratio posteriors rail against the lower mass ratio prior limits}\newnew{, implying that the true probability distribution extends below the lowest value of mass ratio contained within the prior,} \new{and the eccentricity distribution inferred for both events is uninformative. We leave eccentric analyses of systems containing low-mass black holes or neutron stars for future work.}
\end{itemize}

\subsection{Caveats}
Our analysis method leads our results to have the following caveats:
\begin{enumerate}
    \item Since the waveform model that we employ, \texttt{SEOBNRE}, does not support misaligned spins, we are not able to disentangle the effects of orbital eccentricity and spin-induced precession on the signal. This may lead us to infer non-zero eccentricity for a quasi-circular system undergoing spin-precession; see \cite{Romero-Shaw:2020:GW190521}.
    \item \texttt{SEOBNRE} enforces a dimensionless aligned-spin magnitude upper limit of $0.6$. Any binary that is truly highly-spinning will produce a signal that is poorly-specified by our choice of waveform model, and will therefore bias our results.
    \item Similarly, the upper limit of our eccentricity prior is $0.2$ at a detector-frame gravitational-wave frequency of $10$~Hz. Any binary with an eccentricity higher than this will not be correctly specified by our choice of waveform model. However, we have seen in the case of GW190521 that systems consistent with larger eccentricities \citep{JuanHeadOn, Gayathri, Gamba:2021:GW190521} can still show signs of high eccentricity in our analyses, railing against the model-enforced upper limit of the prior \citep[see, e.g.,][]{Romero-Shaw:2020:GW190521}. \newnew{Such railing, in which the peak of the posterior distribution is seen to ``pile up`` at the prior boundary, implies that the truly most probable region of the parameter space likely exists outside the range covered by the prior.}
    \item \new{Waveforms produced with \texttt{IMRPhenomD} and \texttt{SEOBNRE} do not contain higher harmonics, which may mislead inference when higher harmonics are present in the data. Neglecting higher-order modes can bias recovery of parameters like spins and mass ratios \citep{Shaik:2020:NeglectHOM}, so we expect that inferences of eccentricity would also be biased. Although there are no tailored studies to assess the eccentricity bias when higher-order modes are neglected, quasi-circular and spin-precessing waveform templates with higher-order modes may be more likely mistaken for eccentric waveforms than those without higher-order modes \citep{Romero-Shaw:2020:GW190521}. Recent analyses including both eccentricity and higher-order modes \citep{Iglesias:2022:HOM} obtain qualitatively consistent results to previous analyses neglecting higher-order modes.}
    \item \texttt{SEOBNRE} sets the initial argument of periapsis based on the fixed starting frequency of waveform generation. This parameter is therefore not adjustable and cannot be sampled over. We anticipate that being able to sample over this parameter could lead to a shift in the locations of the peaks of the recovered eccentricity distributions in events with high signal-to-noise ratio (SNR $\approx 30$), but the consequences of neglecting this parameter for the events studied here are likely to be small \citep{Clarke:2022:AOP}.
    \item Different waveform models and simulations use different definitions of eccentricity, and use different prescriptions to set initial conditions. This means that an eccentricity inferred with \texttt{SEOBNRE} does not exactly equate to the eccentricity that would be inferred with another model, and that comparisons to predictions made by simulations of dynamical environments should be taken as indicative rather than absolute. Work is ongoing to establish a translation guide between the eccentricities defined by various simulations \citep[][]{Knee:2022:RosettaStone}.
    \item We restrict our analyses to events that have been flagged as likely compact binary merger signals by LVK searches. In order to be flagged as such, a signal must bear significant resemblance to a quasi-circular inspiral track so that it achieves a high match with the waveforms used in those templated searches. As a result, the events we analyse are highly likely to have small or negligible eccentricities.
\end{enumerate}

\section{Eccentricity measurements}\label{sec:results}

\begin{figure*}
    \centering
    \includegraphics[width=\textwidth]{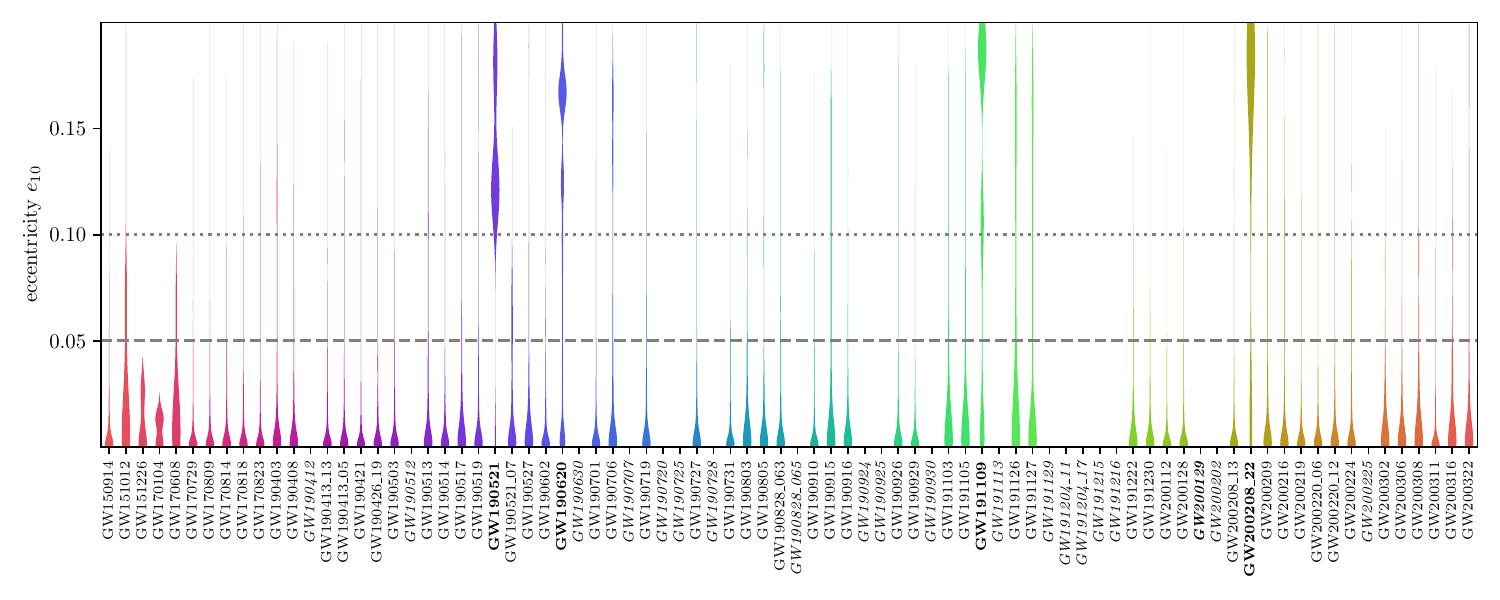}
    \caption{Marginal posterior distributions on $e_{10}$ for all adequately-sampled binary black hole merger events in GWTC-3 are shown in colour. The width of the violin at each value of eccentricity is proportional to the posterior distribution at that value.  The names of undersampled events are italicised and their posteriors are not shown. The names of the four events that show considerable support for $e_{10} \geq 0.05$ are emboldened.}
    \label{fig:eccentricity_violin}
\end{figure*}

\begin{table*}
\centering
\caption{A summary of the detector-frame eccentricity measurements for \new{newly-analysed} events in GWTC-3 with a \new{natural} log (ln) Bayes factor greater than $0$ for the hypothesis that $e_{10} \geq 0.05$ and number of effective samples $\geq 100$.
The two events with greater than $50\%$ of their posterior support at $e_{10} \geq 0.05$ are emboldened.
The second and third columns provide the percentage of posterior support for $e_{10}>0.1$ and $e>0.05$, two values typically used as thresholds for currently detectable binary eccentricity at \unit[10]{Hz}. 
However, these values are somewhat arbitrary, as each individual signal has a different detectable eccentricity threshold.
Furthermore, simulations that quote eccentricity $e_{10} \geq 0.1$ do so in the source frame; one must redshift any detector-frame eccentricity measurements to perform a direct comparison to such simulations (see Section \ref{sec:population}).
The third (fourth) column provides the ln Bayes factor for the hypotheses that $e_{10} \geq 0.1$ ($0.05$) against the hypothesis that $e_{10} < 0.1$ ($0.05$).
The final column states the number of effective samples in the eccentric posterior after reweighting.
The quantities presented in this table are provided for \new{newly-studied} events in GWTC-3 without significant support for eccentricity in the Appendix in Table \ref{tab:results-not-eccentric}.
\label{tab:results-eccentric}}
\bgroup
\def\arraystretch{1.5}
\begin{tabular}{c|c|c|c|c|c}
Event name & $e_{10} \geq 0.1$ (\%) & $e_{10} \geq 0.05$ (\%) & $\ln \mathcal{B} (e_{10} \geq 0.1)$  & $\ln \mathcal{B} (e_{10} \geq 0.05)$ & $n_{\rm eff}$ \\
\hline
GW190403 & $16.64$ & $27.53$ & $0.83$ & $0.56$ & $6358$ \\
GW190805 & $14.33$ & $23.78$ & $0.19$ & $0.03$ & $719$ \\
GW191105 & $10.58$ & $18.52$ & $0.29$ & $0.12$ & $718$ \\
\textbf{GW191109} & $\mathbf{62.52}$ & $\mathbf{71.53}$ & $\mathbf{1.89}$ & $\mathbf{1.49}$ & $\mathbf{7125}$ \\
GW191126 & $26.28$ & $33.79$ & $1.27$ & $0.92$ & $293$ \\
GW191127 & $23.68$ & $33.33$ & $1.20$ & $0.88$ & $436$ \\
\textbf{GW200208\_22} & $\mathbf{70.78}$ & $\mathbf{73.06}$ & $\mathbf{1.94}$ & $\mathbf{1.39}$ & $\mathbf{219}$ \\
GW200209 & $16.58$ & $26.63$ & $0.64$ & $0.44$ & $43848$ \\
GW200216 & $10.62$ & $22.43$ & $0.26$ & $0.25$ & $3108$ \\
GW200322 & $12.57$ & $21.37$ & $0.12$ & $0.07$ & $10203$ \\
\end{tabular}
\egroup
\end{table*}

\begin{figure*}
    \centering
    \includegraphics[width=0.6\textwidth]{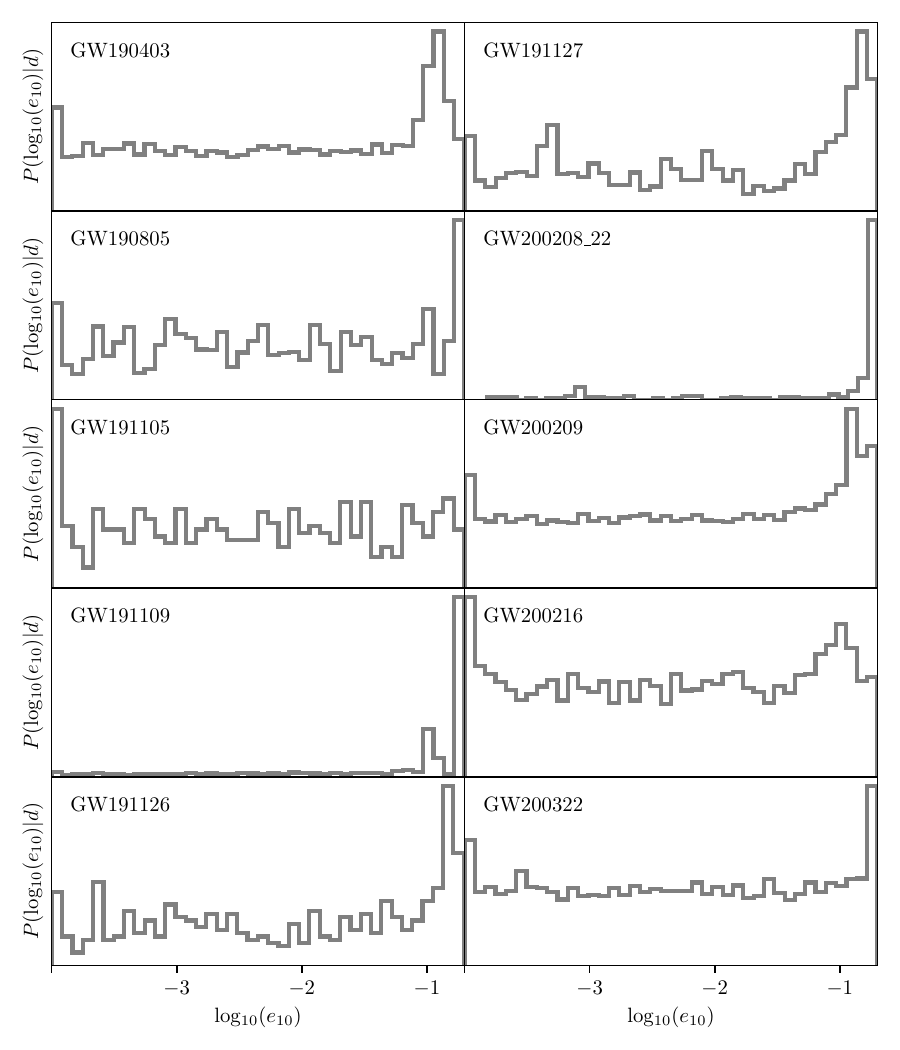}
    \caption{Marginal posterior distributions on log$_{10}(e_{10})$ for events highlighted in Table \ref{tab:results-eccentric}, which have a significant fraction of their posterior support above $e_{10}=0.05$ and a positive ln Bayes factor when comparing the eccentric hypothesis to the quasi-circular hypothesis. We label each panel with the name of the event.}
    \label{fig:eccentric_posteriors_1D}
\end{figure*}

In Figure \ref{fig:eccentricity_violin}, we provide marginal one-dimensional eccentricity posterior distributions for \Nevents~BBH candidates.
We note that we have removed two events that were found to have below-threshold significance in GWTC-2.1 \citep[GW190924A and GW190909A;][]{GWTC-2-1}, from this plot and from all analyses presented in Section \ref{sec:population}.
\new{In Table \ref{tab:results-eccentric}, we provide numerical summaries of the eccentricity measurements for newly-analysed events in GWTC-3, including ln Bayes factors for the \textit{detectably}-eccentric region of the parameter space relative to the quasi-circular hypothesis. 
To calculate these values, we restrict the eccentricity posteriors to $e_{10} \geq 0.05$, the minimum expected detectable eccentricity \citep{Lower18}, and $e_{10} \geq 0.1$, since this is an often-used conservative threshold for detectable eccentricity resulting from dynamical binary formation \citep[e.g.,][]{Wen02, Gondan17, Samsing18, Rodriguez18a, Rodriguez18b, GondanKocsis2021}.
We restrict the eccentricity prior to astrophysically-motivated ranges for these Bayes-factor calculations to avoid contamination from quasi-circular samples, which do not represent the eccentric hypothesis. }
We present in Figure \ref{fig:eccentric_posteriors_1D} marginal posteriors on log$_{10}(e_{10})$ for the ten \new{newly-analysed} events in GWTC-3 that have positive ln Bayes factors in favour of the eccentric hypothesis when compared to the quasi-circular hypothesis.
\new{We adopt the convention that a ``detection'' of eccentricity is not made unless ln~$B \geq 8$, so do not claim that any of these events definitively prefer the eccentric model over the quasi-circular model.}
Posterior probability distributions on all parameters of all analysed events are provided online.\footnote{\href{https://github.com/IsobelMarguarethe/eccentric-GWTC-3}{github.com/IsobelMarguarethe/eccentric-GWTC-3}}

\subsection{\new{New events in} GWTC-3 with majority posterior support for $e_{10} \geq 0.05$}

\begin{figure}
    \centering
    \includegraphics[width=0.45\textwidth]{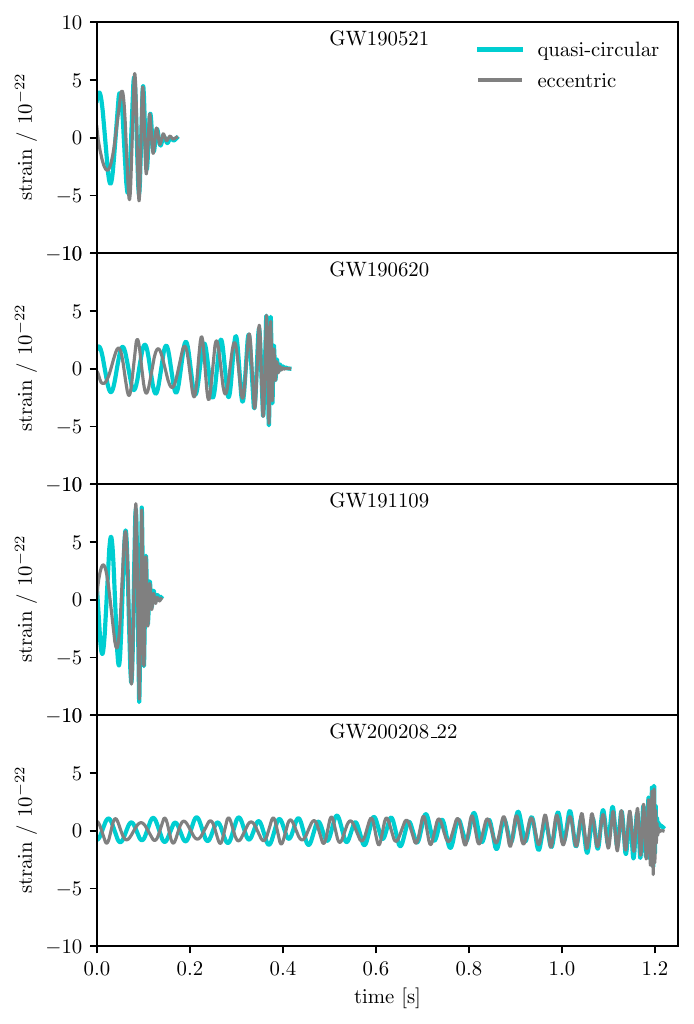}
    \caption{Time-domain waveforms for the median posterior parameters inferred for the \Necc~events with support for $e_{10} \geq 0.05$ observed by the LIGO and Virgo detectors so far. The preferred \texttt{SEOBNRE} waveforms are shown in grey and the corresponding quasi-circular \texttt{IMRPhenomD} waveforms are shown in teal. The panel width spans $1.25$ seconds.}
    \label{fig:eccentric-waveforms}
\end{figure}

There are two new events that show significant evidence for eccentricities above $e_{10} = 0.05$ \new{within their posterior probability distributions}, with $\geq 50\%$ of their posterior probability support above $e_{10} = 0.05$.
For these events, and for the two eccentric candidates GW190521 and GW190620 \citep{Romero-Shaw:2020:GW190521, Romero-Shaw:2021:GWTC-2-ecc}, we conduct our analysis with more aggressive sampler settings (4000 live points and 200 walks) to obtain a higher number of effective samples. These results are those shown in Fig. \ref{fig:eccentricity_violin} and all other figures, and in Table \ref{tab:results-eccentric}.
The waveforms corresponding to the median posterior parameters for GW190521, GW190620, GW191109 and GW200208 are shown in Fig. \ref{fig:eccentric-waveforms}.

\subsubsection{GW191109}

\begin{figure*}
    \centering
    \includegraphics[width=0.45\textwidth]{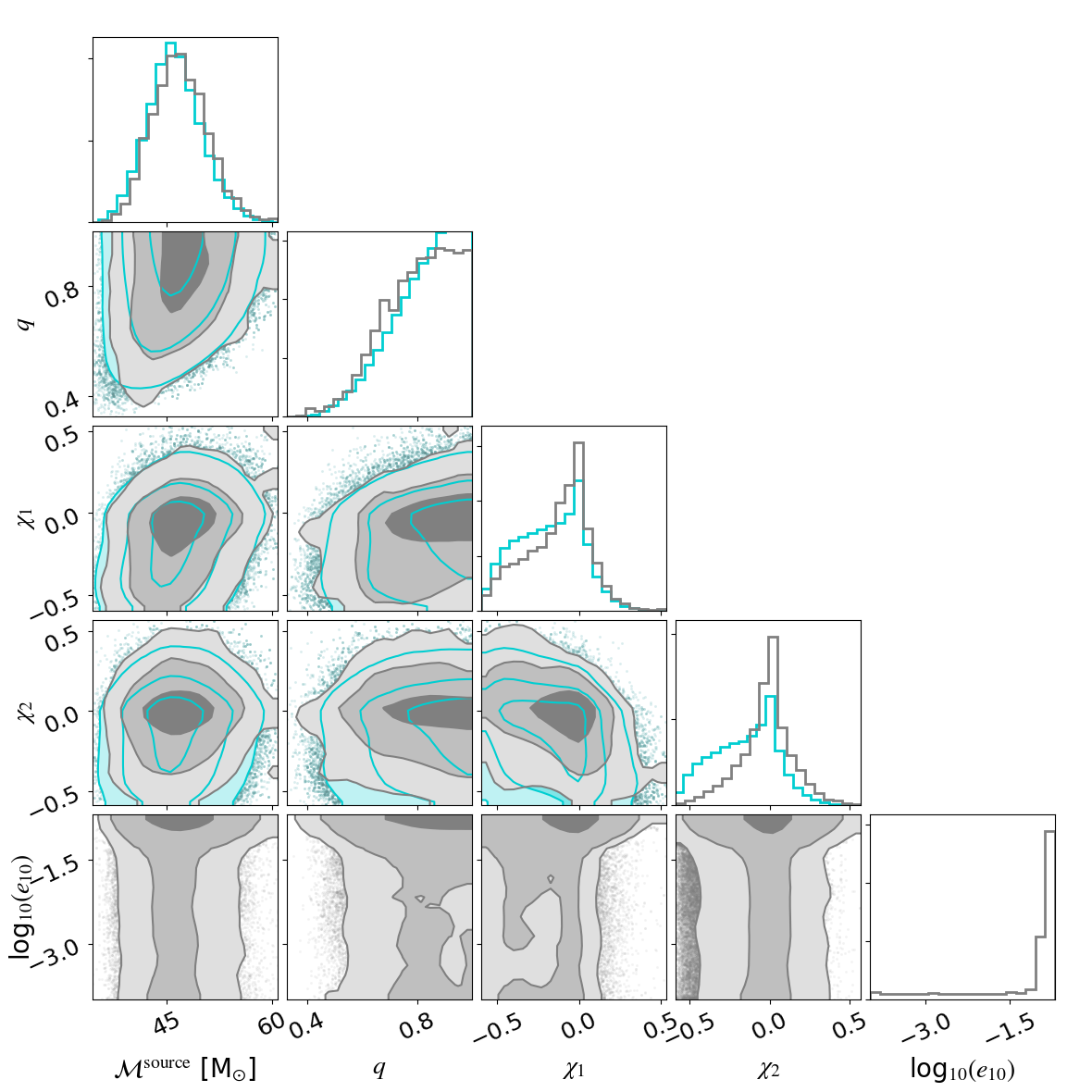}
    \includegraphics[width=0.45\textwidth]{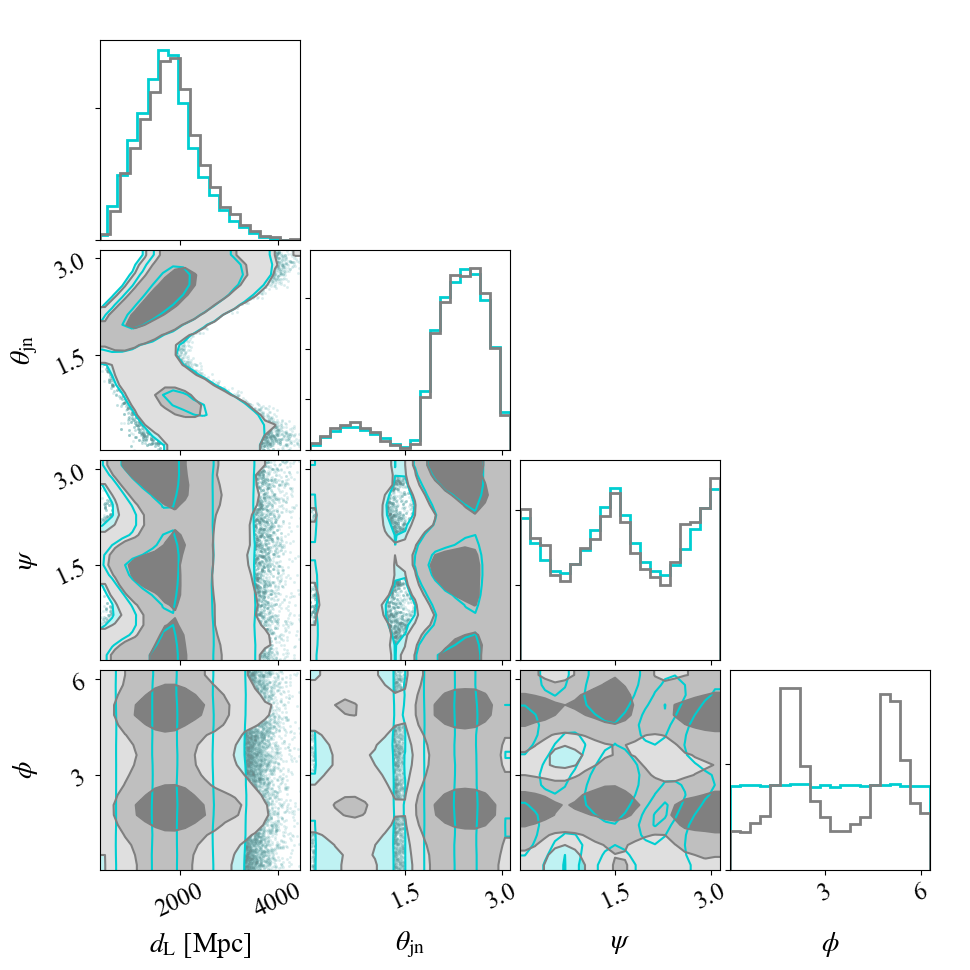}
     \caption{Intrinsic (left) and extrinsic (right) parameters inferred for GW191109. Posterior probability distributions under the eccentric model are shown in grey, while the underlying quasi-circular posteriors used to importance-sample the parameter space are shown in teal. 
    \label{fig:GW191109A}}
\end{figure*}

GW191109 was found by the LVK analysis to have the highest support for negative spin of all \new{new} GWTC-3 events.
Our initial quasi-circular analysis recovers this preference for negatively-aligned spins, but when we reweight to the eccentric posterior, higher spin magnitudes have lower weights.
This is similar to the reweighting behaviour observed for eccentric candidate GW190620 \citep{Romero-Shaw:2021:GWTC-2-ecc}, although we retain some appreciable deviation from the spin prior for $\chi_1$ after reweighting for GW191109.
Other parameters are consistent with those recovered in the LVK analysis.
Proposal (quasi-circular) posteriors for GW191109 are shown in Fig. \ref{fig:GW191109A} in teal, and target (eccentric) posteriors are overplotted in grey.

We find that GW191109 has $72.19$\% of its posterior support above $e_{10} = 0.05$, $62.63$\% of its posterior above $e_{10}=0.1$, and a ln Bayes factor of ln~$\mathcal{B} = 1.49$ ($1.89$) in favour of the $e_{10} \geq 0.05$ ($e_{10} \geq 0.1$) hypothesis relative to the quasi-circular hypothesis.
For this event, we obtain a reweighting efficiency of $4.52\%$, and obtain $n_{\rm eff} = 7125$.
Visible in Figure \ref{fig:eccentricity_violin} is the double-peaked structure of the eccentricity posterior for GW191109: the main peak is at the upper limit of the eccentricity prior, $e_{10}=0.2$, but there is a subdominant mode at $e_{10}\approx0.1$.

The eccentric (target) posterior distribution on the phase of coalescence, $\phi$, for GW191109 has periodic peaks that are not present in the quasi-circular (proposal) posterior distribution.
Since GW191109 is relatively high-mass, its $e_{10}$ measurement gives the shape of its orbit just a few cycles before the coalescence itself.
Having distinctly non-zero orbital eccentricity at this point means that the orbit close-to-merger is elongated, so the gravitational-wave emission varies more strongly with $\phi$.
\newnew{Additionally, $\phi$ is likely to closely correlate with the argument of periapsis, $\omega$; while $\phi$ describes the phase of the orbit, $\omega$ describes the angle of rotation of the orbit itself.
Since $\omega$ is set indirectly through the reference frequency $f_{\rm ref}$ and eccentricity $e_{10}$ in our analyses, the narrow range of $e_{10}$ and fixed $f_{\rm ref}$ may restrict $\omega$ enough that only certain values of $\phi$ can produce waveforms consistent with the data.}

\subsubsection{GW200208\_22}

\begin{figure*}
    \centering
    \includegraphics[width=0.45\textwidth]{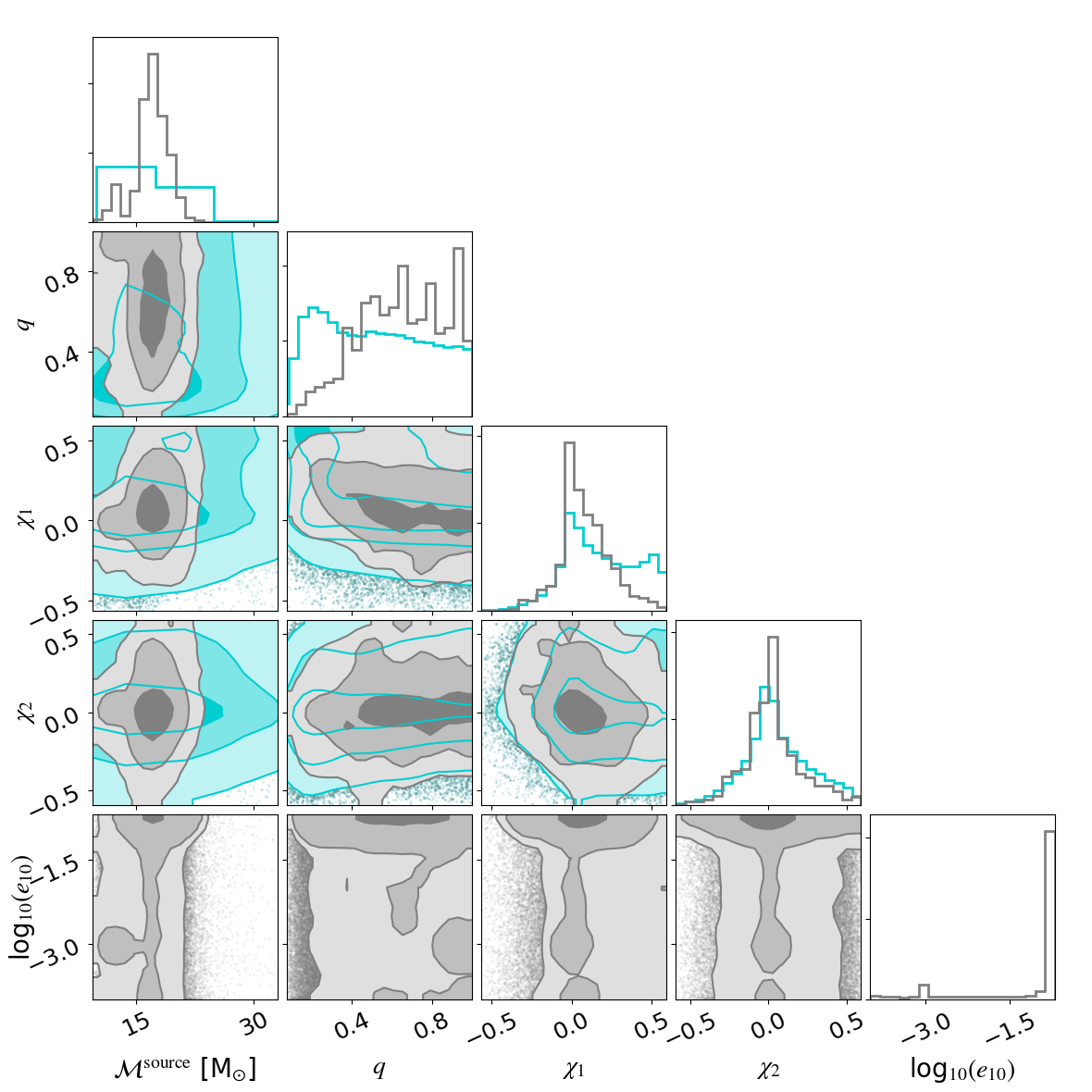}
    \includegraphics[width=0.45\textwidth]{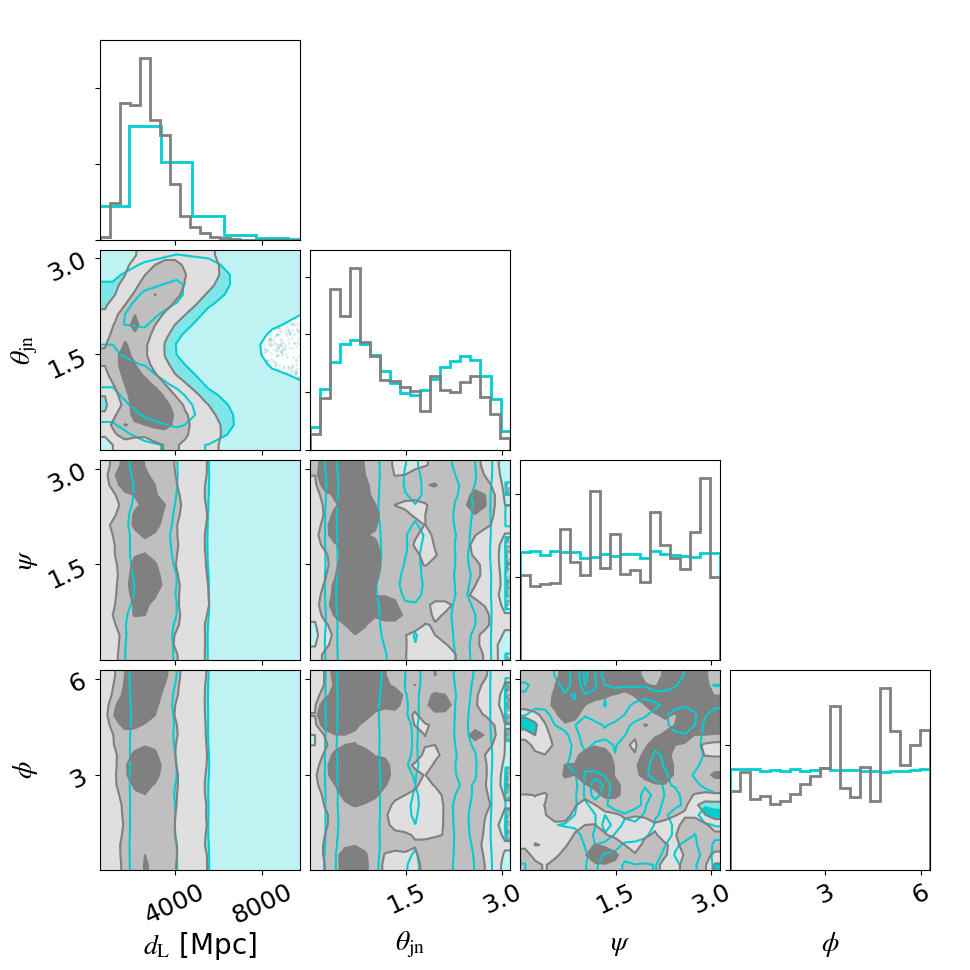}
    \caption{Intrinsic (left) and extrinsic (right) parameters inferred for GW200208\_22.  
    \label{fig:GW200208B}}
\end{figure*}

We find that, when analysed with our default sampling settings, the mass ratio posterior for GW200208\_22 rails against the lower limit of the prior\newnew{, indicating that the lower limit is too high to capture the full extent of the posterior}. Therefore, for this event, we reduce the lower mass ratio prior limit from 0.125 to 0.025. We notice also that the posterior rails against the upper limits of $\chi_1$ and $\chi_2$ at $0.6$\newnew{, implying that the true distribution extends above the upper limits allowed by the prior}. However, since this is a limit enforced by \texttt{SEOBNRE}, we do not relax this limit.
In Fig. \ref{fig:GW200208B}, we show proposal (quasi-circular) posteriors for GW200208\_22 in teal and target (eccentric) posteriors in grey.

We find that GW200208\_22 has $76.71$\% of its posterior support above $e_{10}=0.05$, with $73.52$\% above $e_{10}=0.1$, and ln Bayes factor for $e_{10} \geq 0.05$ ($e_{10} \geq 0.1$) relative to the quasi-circular hypothesis of $1.45$ ($1.92$).
In our analysis, GW200208\_22 has a a source-frame chirp mass of $17.38^{+3.10}_{ -4.54}$~M$_\odot$.
The reweighting efficiency for GW200208\_22 is relatively low at $0.18$\%, consistent with our expectations for low-chirp mass (long-duration) eccentric signals.
Additionally, waveform model choice been shown to be important for this event \citep{GWTC-3}, so posterior differences that are not correlated with eccentricity may be a result of waveform systematics.
We obtain $219$ effective samples for this event. 

Our posteriors peak at a lower total mass ($\sim 41$ M$_\odot$) than that found in the LVK analysis ($\sim 63$ M$_\odot$), although the posteriors do overlap: while the LVK posteriors and our \texttt{IMRPhenomD} posteriors are multimodal, the eccentric posterior favours the lower-mass peak. 
The median primary mass recovered with the eccentric model ($\sim 25$~M$_\odot$) is significantly lighter than the median of the LVK analysis ($\sim 51$~M$_\odot$), for the same reason. 
As a result, the median luminosity distance recovered is roughly $1$~Gpc smaller than the median LVK result. 
It is possible for eccentric systems to masquerade as higher-mass quasi-circular systems when their gravitational-wave signals are analysed assuming quasi-circularity, since eccentricity can drive a binary to merge on a faster timescale and merge at a lower frequency \citep[see, e.g.,][]{JuanHeadOn, Favata:2022:eccentricity-constraints}.
In addition to the inclusion of eccentricity, another reason for the discrepancy in median posterior parameters may be our enforced limit on the spin prior magnitude due to the limitations of \texttt{SEOBNRE}: while the LVK analysis finds that $\chi_1 \geq 0.29$ ($90\%$ credibility) with $51\%$ of its posterior support above $\chi_1 = 0.8$, we infer a posterior that rails against the upper limit of the prior at $\chi_1 = 0.6$. 

As in the LVK analysis, the mass ratio posterior recovered with \texttt{IMRPhenomD} is quite flat, with a slight preference for unequal masses ($m_2 / m_1 \approx 0.3$).
When we reweight to \texttt{SEOBNRE}, the long tail out to high masses is downweighted and only the peak at chirp mass $\mathcal{M} \approx 16$ M$_\odot$ remains. More equal mass ratios are favoured by the eccentric model over unequal masses. Higher values of $\chi_1$ and $\chi_2$ are also downweighted, with the eccentric model preferring samples with high values of eccentricity and low values of spin. Again, the reason for the difference may be the enforced spin prior limitation. It is possible that the data is best-fit by a waveform with $\chi_1 \sim 0.9$ and $e_{10} = 0$, but prefers a waveform with $\chi_1 = 0$ and $e_{10} = 0.2$ to one with $\chi_1 \sim 0.5$. In this case, because of our restricted prior, the vast majority of the eccentric posterior is at high eccentricities with low spins. Previous analyses have show that, when spin amplitude is restricted, higher eccentricities may be favoured \citep[see][]{OSheaKumar2021}

\subsection{Other \new{new} events in GWTC-3 with non-negligible support for $e_{10} \geq 0.05$}

Another two candidates from O3b, GW191126 and GW191127, have $> 30\%$ of their posterior support at $e_{10} \geq 0.05$ and ln $\mathcal{B} (e_{10} \geq 0.1) >1.0$.
Their marginal eccentricity posterior distributions show peaks at our prior upper limit of $e_{10}=0.2$ and non-negligible tails down to lower eccentricities. 
While we also analyse these event with more aggressive sampler settings, we do not discuss these events in detail here; we reserve a detailed analysis of these events, in conjunction with the marginal eccentric candidates presented in \citet{Romero-Shaw:2021:GWTC-2-ecc}, for future work. 

\subsection{Notable \new{new} events in GWTC-3 with $e_{10} \leq 0.05$}

\subsubsection{Mass-gap events}
GW190426\_19 has the highest mass of all binary mergers reported by the LVK, with both components more massive than predicted by isolated evolution: $m_1 = 106.9^{+41.6}_{-25.2}$ M$_\odot$, $m_2 = 76.6^{+26.2}_{-33.6}$ M$_\odot$ \citep{GWTC-2-1}. Our analysis of GW190426\_19 recovers parameters consistent with those recovered in the LVK analysis, including the slight deviation from the prior at higher values of $\chi_{\rm eff}$. 
Another high-mass event, GW200220\_06, is found by the LVK to have mass-gap components: $m_1 = 87^{+40}_{-23}$ M$_\odot$, $m_2 = 61^{+26}_{-25}$ M$_\odot$ \citep{GWTC-3}, consistent with our findings.
We find that GW190426\_19 and GW200220\_06 do \textit{not} contain hints of orbital eccentricity. 

GW190426\_19 and GW200220\_06 have high reweighting efficiencies of $72\%$ and $83\%$ respectively, with $11\%$ and $13\%$ of their posterior support above $e_{10} = 0.05$. 
Reweighting to \texttt{SEOBNRE} pushes the preferred mass-ratio and source-frame chirp mass to slightly lower values for GW190426B, but this shift does not appear correlated with eccentricity; we put the difference down to waveform systematics.
There is virtually no difference between the quasi-circular and eccentric posteriors for GW200220\_06.

Being consistent with quasi-circular at $10$~Hz does not mean that these binaries are not dynamically formed: we expect only $\sim 4\%$ of our detected mergers from globular clusters to retain detectable eccentricity at this frequency \citep{Zevin:2021:seleccentricity}, and more massive mergers circularise at lower frequencies than their lower-mass counterparts.
It is nonetheless worth noting that the eccentricity, spin magnitude, and spin-tilt measurements for these systems are inconclusive.
If they contained merger remnants, which became bound through dynamical interactions, their dimensionless spin magnitudes should be $\chi_i \sim 0.7$, and their spin tilt angles would likely be misaligned \citep[e.g.,][]{Pretorius:FinalSpin:2005, Gonzalez:FinalSpin:2007, Buonanno:FinalSpin:2008}.
Alternatively, such massive binaries may form in isolation if the pair-instability mass gap is narrower than predicted \citep[as suggested in the wake of GW190521 by, e.g.,][]{Costa:2021:0521}, or if our standard priors on mass ratio are misleading inference \citep[also suggested to explain GW190521 by][]{Fishbach:2020:0521}.

\subsubsection{Spinning events}

A number of the events that strongly support nonzero spins are undersampled after the reweighting process: GW200129, which exhibits support for signs of spin-induced precession, GW191204, which has a $\chi_{\rm eff}$ posterior tightly constrained away from zero, and GW191216, which has negligible support for $\chi_{\rm eff} = 0$ \citep{GWTC-3}.
However, some are adequately-sampled: GW191103, for example, which has $175$ samples after reweighting.
The marginal eccentricity posterior for this event is uninformative, and correlated with spin: lower magnitudes of $\chi_{1}$ are favoured for samples with $e_{10} \geq 0.05$.
The eccentric reweighting process disfavours larger values of $\chi_{1}$, while the marginal $\chi_{2}$ posterior is relatively unchanged.
GW191103A has only $17\%$ of its posterior support above $e_{10} = 0.05$ after reweighting.

\subsection{Spin-induced precession or eccentricity?}\label{subsec:eccprecc}

There are currently no waveform models that incorporate the simultaneous effects of eccentricity and spin-induced precession on the signal.
Since the two effects can cause similar phase and amplitude modulations in gravitational-wave signals \citep[e.g.,][]{JuanHeadOn}, they can cause spin-aligned analyses to recover eccentricity, or quasi-circular analyses to recover misaligned spins \citep{JuanBosonStars, Romero-Shaw:2020:GW190521}.
Therefore, any non-zero eccentricity measurements that we infer in our analysis may, in actuality, be caused by the binary having misaligned spins.

Whilst we cannot simultaneously infer the presence of spin-induced precession and eccentricity, we can attempt to deduce which effect is more likely to be present in the signal.
We perform analyses on GW190521, GW190620, GW191109 and GW200208\_22 using a precessing waveform approximant; see \citet{Romero-Shaw:2020:GW190521} for an extensive comparison between the eccentric and spin-precessing hypotheses for GW190521.
For the study presented here, we employ one of the preferred waveforms used in the transient catalogue of the LVK \citep{GWTC-2, GWTC-2-1, GWTC-3}: quasi-circular spin-precessing model \texttt{IMRPhenomXPHM} \citep{Pratten:2021:PhenomXPHM}, allowing the full range of available spin orientations in our priors and component spin magnitudes up to $0.89$, and using the same aggressive sampler settings as employed for our follow-up analyses using \texttt{SEOBNRE}.

\begin{table}
    \centering
    \begin{tabular}{c|c|c}
         Event & ln~$\mathcal{B}_{E/P} (e_{10} \geq 0.1)$ & ln~$\mathcal{B}_{E/P} (e_{10} \geq 0.05)$ \\ 
         \hline
        GW190521 & 3.06 & 2.30 \\ 
        GW190620 & 2.52 & 2.10 \\ 
        GW191109 & -1.74 & -2.14 \\ 
        GW200208\_22 & 1.51 & 0.96 \\ 
    \end{tabular}
    \caption{Natural log (ln) Bayes factors for the eccentric, spin-aligned hypothesis calculated with waveform model \texttt{SEOBNRE} compared to the quasi-circular, spin-precessing hypothesis calculated with waveform model \texttt{IMRPhenomXPHM} for three potentially-eccentric sources (for an extensive analysis of GW190521 with other spin-precessing approximants, see \citet{Romero-Shaw:2020:GW190521}). We compare hypotheses for astrophysically-motivated regions of the prior space: above $e_{10}=0.1$, a common threshold given in the literature, and above $e_{10}=0.05$, the approximate sensitivity limit for current detectors \citep{Lower18, Romero-Shaw:2019:GWTC-1-ecc, Romero-Shaw:2021:GWTC-2-ecc}. \new{In no event is the eccentric hypothesis strongly preferred over the precession hypothesis, or vice versa, using the convention that one argument is strongly preferred over another if $|$ln~$B| \geq 8$}.}
    \label{tab:spin_precessing_bayes_factors}
\end{table}

Table \ref{tab:spin_precessing_bayes_factors} contains the relative natural log (ln) Bayes factors of the spin-aligned, eccentric hypothesis (calculated from our \texttt{SEOBNRE} posteriors above thresholds of $e_{10}=0.05$ and $0.1$) against the quasi-circular, spin-precessing hypothesis (caculated from our \texttt{IMRPhenomXPHM} posteriors) for GW190521, GW190620, GW191109 and GW200208\_22. There is a marginal preference for the eccentric hypothesis for GW190521, GW190620, and GW200208\_22, and a marginal preference for the spin-precessing hypothesis for GW191109; however, in no case is the evidence for either hypothesis overwhelming. Thus, each one of our potentially-eccentric candidates \textit{could} be eccentric, spin-precessing, or both.

\section{Eccentricity in the population}\label{sec:population}

\begin{figure}
    \centering
    \includegraphics[width=0.45\textwidth]{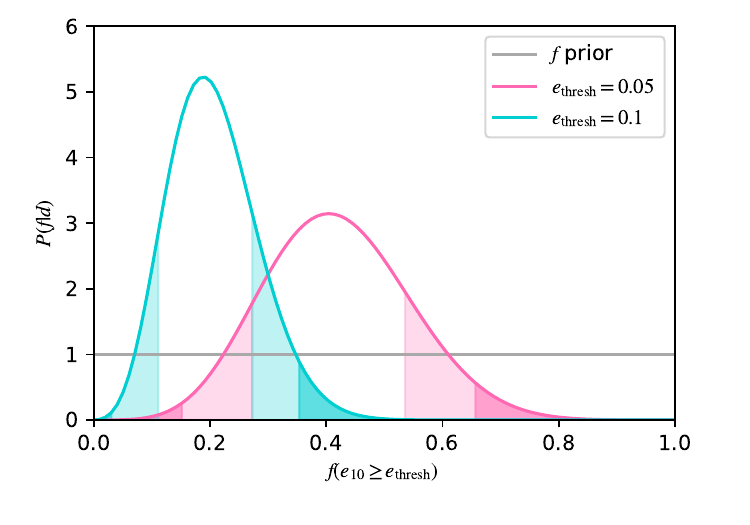}
    \caption{Posterior probability distributions for the fraction of eccentric mergers above $e_{\rm thresh} = 0.05$ and $0.1$ at $10$~Hz. The distributions peak at $0.19$ (teal curve, representing $e_{\rm thresh} = 0.1$) and $0.40$ (pink curve, representing $e_{\rm thresh} = 0.05$). The uniform $f$ prior is indicated with a grey horizontal line. Unfilled, lightly-filled and darkest-filled areas represent the symmetric $1\sigma$, $2\sigma$ and $3\sigma$ credible ranges, respectively. The $f(e_{10}\geq0.05)=0$ and $f(e_{10}\geq0.1)=0$ points are excluded at greater than $2\sigma$. The $f(e_{10}\geq0.05)=0$ point is outside of the symmetric $3\sigma$ credible region. 
    }
    \label{fig:population_fraction}
\end{figure}

As the catalogue of mergers grows, it becomes increasingly likely that random noise fluctuations emulate the effects of eccentricity in a subset of BBH merger signals.
Additionally, while we highlight four events that have clear peaks above $e_{10}=0.05$ in their eccentricity posterior probability distributions, there are multiple other events that show significant support for $e_{10}\geq0.05$.
We wish to quantify the fraction of observed mergers that truly support the eccentric merger hypothesis, without assuming any specific formation channel.
We perform population analyses under the hypothesis that some binaries have some support for eccentricity $e_{10}$ above some threshold eccentricity, $e_{\rm thresh}$.
We consider two possibilities: that $e_{\rm thresh} = 0.05$ and that any eccentricity lower than this is not detectable; and a more conservative hypothesis with $e_{\rm thresh} =  0.1$.
We calculate a likelihood for $f$, the fraction of the population support for $e_{10} \geq e_{\rm thresh}$:

\begin{align}
    {\cal L}(d | f) = \prod_k
    \Bigg( &
    f \int_{e_{\rm thresh}}^{e_{\rm max}} de \, \pi(e) {\cal L}(d_k | e)
    \nonumber \\ & +
    (1-f) \int_{e_{\rm min}}^{e_{\rm thresh}} de \, \pi(e) {\cal L}(d_k | e)
    \Bigg).
\end{align}    
Here, $k$ represents each event in our population, $e_{\rm min}$ and $e_{\rm max}$ are our eccentricity prior bounds, and $\pi(e) {\cal L}(d_k | e)$ is the marginal posterior probability distribution for the eccentricity of event $k$. 
Drawing proposals for the value of $f$ from a uniform prior and computing ${\cal L}(d | f)$ over this range produces a posterior probability distribution for $f$. This posterior is plotted in Figure \ref{fig:population_fraction} for both $e_{\rm thresh}$ conditions.

The highest-probability $f$ representing the fraction of observed BBH with $e_{10} \geq 0.05$ is $0.40$, corresponding to $25$ mergers, while the maximum-posterior $f$ for BBH observed with $e_{10} \geq 0.1$ is $0.19$, corresponding to $12$ mergers.
We can exclude $f=0$ with greater than $2\sigma$ credibility in both cases.
We therefore conclude that the population support for eccentricity in GWTC-3 is consistent with a non-negligible fraction of mergers exhibiting detectable eccentricity at $10$~Hz.

\new{The fact that we include only the adequately-sampled events in the calculation of $f$ may bias our results.
Reweighting can fail to produce sufficient samples if the target (eccentric) posterior probability distribution exists in a region of the parameter space not well-covered by the proposal (quasi-circular) posterior.
For example, injection studies \citep{OSheaKumar2021} show that a quasi-circular waveform used to recover an eccentric signal can result in a posterior distribution that does not contain the injected mass parameters.
In such a case, an analysis reliant on likelihood reweighting is likely to either be undersampled or return an eccentricity posterior distribution consistent with zero eccentricity.}

\newnew{Because we use a reweighting-based method and neglect undersampled events---both of which bias us towards detecting more events consistent with quasi-circularity---our estimates of $f$ can be considered as lower limits when we assume that none of the events are spin-precessing. 
On the other hand, if the potential degeneracy between eccentricity and precession \citep[e.g.,][]{Romero-Shaw:2020:GW190521, JuanHeadOn} is leading quasi-circular systems to be interpreted as eccentric by our analysis, then our inferred values of $f$ may be artificially high.
More accurate estimates of $f$ require studies to determine exactly which regions of the parameter space undersampling and eccentricity-precession degeneracy are most prevalent, and are left for future work.}

\subsection{Implications for population formation channels}
\label{sec:formation-implications}

\begin{figure*}
    \centering
    \includegraphics[width=0.45\textwidth]{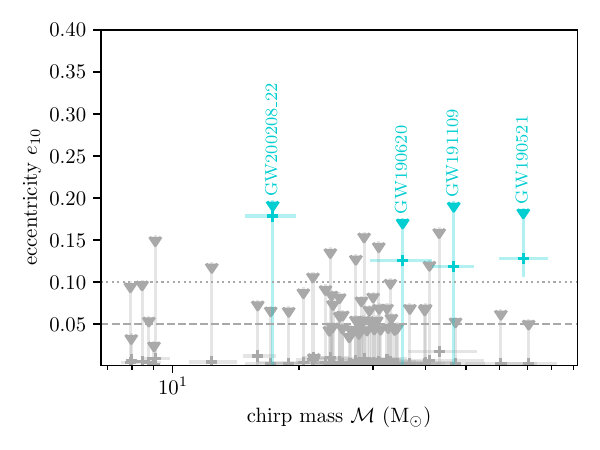}
    \includegraphics[width=0.45\textwidth]{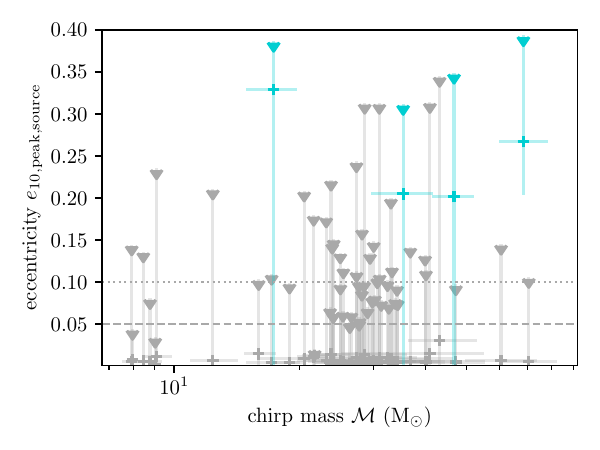}
    \caption{The upper $90\%$ credible limits for all BBH candidates in GWTC-3 are shown as downward-facing triangles, plotted against their median source-frame chirp mass. The panel on the left shows eccentricities as measured at a Keplerian gravitational-wave frequency of $10$~Hz in the detector frame, while the right-hand panel shows eccentricities shifted to a peak gravitational-wave frequency of $10$~Hz in the source frame. The symmetric $90\%$ credible chirp mass range is indicated for each binary by a horizontal line through their median eccentricity. The median eccentricity is plotted against median source-frame chirp mass with a plus symbol. The four events with more than $50\%$ of their posterior probability distribution above $e_{10} = 0.05$ are highlighted in teal and labelled with their names in the left-hand plot, while all other events are shown in grey. Undersampled events are not plotted.
        \label{fig:mass_vs_ecc}}
\end{figure*}

Simulations suggest that $5$--$10\%$ of BBH mergers in dense star clusters should enter the LVK sensitivity band with $e_{10} \geq 0.05$ \citep[see, e.g., ][]{Samsing17,Samsing18,Zevin18,Rodriguez18a,Rodriguez18b, Rodriguez19,Kyle_data_paper}. \new{As noted above,} \citet{Zevin:2021:seleccentricity} showed that, from a simulated population of mergers from the \texttt{CMC Cluster Catalog} \citep{Kyle_data_paper}, only $3.9\%$ of globular cluster mergers could be detected with $e_{10} \geq 0.05$\new{, assuming a templated search using quasi-circular waveforms with otherwise identical parameters to the injected eccentric signals}.
\new{Under this assumption, we expect to detect only $56\%$ of sources with $e_{10} \geq 0.05$, since gravitational-wave signals from compact binary mergers are detected using search methods that assume quasi-circular inspirals.}
\new{This fraction may be higher, if eccentric signals achieve maximum SNR against quasi-circular templates with different source parameters (as suggested by the results of, e.g., \citet{OSheaKumar2021}).
However, if the parameters preferred by the quasi-circular model are totally removed from the true quasi-circular parameters, then the eccentricity posterior obtained through the reweighting method is unlikely to both recover a high eccentricity and be well-sampled; therefore, we use $3.9\%$ as our expected eccentric fraction.}
\new{The predictions in \citet{Zevin:2021:seleccentricity} were obtained} using a different eccentric waveform model, \texttt{TEOBResumS} \citep{Nagar:2018zoe}\new{; for an overview of key differences between \texttt{TEOBResumS} and \texttt{SEOBNRE}, see \citet{Knee:2022:RosettaStone}}.
\new{Since the overlap between \texttt{SEOBNRE} and \texttt{TEOBResumS} is $\mathcal{O} > 90\%$ in LVK noise over the eccentricity range studied here \citep{Knee:2022:RosettaStone}, we assume in this paper that the results of \citet{Zevin:2021:seleccentricity} are robust to waveform choice. 
However, we caution that this may not be the case; a repeated study using \texttt{SEOBNRE} is required to obtain waveform model-specific predictions. 
We reserve this for future work, and note that our distributions for $B_c$ based on predictions using the alternative waveform model are therefore approximate.}

\subsubsection{\newnew{Comparing detector-frame eccentricity measurements to simulation predictions}}

Eccentricity distributions obtained from simulations of dense star clusters are naturally quoted at a reference frequency of $10$~Hz in the source frame, while we measure eccentricity at a reference frequency of $10$~Hz in the detector frame. 
Since redshifting pushes detector-frame frequencies lower than their source-frame origins, this means that the lower limits that we report for possibly eccentric events are overly conservative for binaries in the source frame. 
On the flip side, upper limits reported for non-eccentric events are less conservative for source-frame binaries.
We convert measurements of eccentricity into the source frame by establishing the source-frame frequency corresponding to a detector-frame frequency of $10$~Hz: $f_{\rm source} = 10 (1 + z)$~Hz.
We then back-evolve $e_{10}$ from $f_{\rm source} = 10 (1 + z)$~Hz to $f_{\rm source} = 10$~Hz using Peter's equations \citep{Peters64}.

An additional complication comes from conflicting definitions of the reference frequency at which eccentricity is quoted. 
The reference frequency of the \texttt{SEOBNRE} model, $f_{\rm SEOBNRE}$, is defined relative to a closed Keplerian orbit, with a semi-major axis that changes as the binary inspirals.
Meanwhile, simulations of cluster mergers report eccentricities defined at the peak frequency of gravitational-wave emission, $f_{\rm peak}$. 
Within \texttt{SEOBNRE}, an eccentric correction is applied such that the minimum frequency of gravitational-wave content is $f_{\rm SEOBNRE, min} = f_{\rm SEOBNRE} / (1 - e_{f_{\rm SEOBNRE}}^2)^{1.5}$. 
This means that at $f_{\rm SEOBNRE} = 10$~Hz, the maximum difference between $f_{\rm SEOBNRE}$ and $f_{\rm SEOBNRE, min}$ is $0.63$~Hz (for $e_{f_{\rm SEOBNRE}} = 0.2$). 
We start analysis for most events from $20$~Hz, with the exception of GW190521, which we start from $11$~Hz. 
Therefore, this internal eccentric correction does not cause the waveform to start within our analysis band.
However, our reported $e_{10}$ measurements are not directly comparable to the predictions of globular cluster simulations, since \citep{Wen02}
\begin{equation}
\label{eq:fpeak}
f_{\rm peak} = f_{\rm SEOBNRE} (1 - e_{f_{\rm SEOBNRE}}^2)^{-1.5} (1 + e_{f_{\rm SEOBNRE}})^{1.1954} .
\end{equation}
The maximal difference between $f_{\rm SEOBNRE}$ and $f_{\rm peak}$ is therefore $3.2$~Hz (for $e_{f_{\rm SEOBNRE}} = 0.2$).

We convert from an eccentricity distribution defined at $f_{\rm SEOBNRE} = 10$~Hz in the detector frame to its equivalent at $f_{\rm peak} = 10$~Hz in the source frame using Eq. \ref{eq:fpeak} and Peter's equations \citep{Peters64}.
We caution that these results be taken as indicative rather than exact measurements: evolving back to $f_{\rm peak} = 10$~Hz pushes $f_{\rm SEOBNRE}$ to lower frequencies.
Therefore, we are implicitly assuming that the inspiral of the system was unperturbed at lower frequencies, corresponding to earlier times.

In the left-hand panel of Fig. \ref{fig:mass_vs_ecc}, we plot measured median and upper $90\%$ credible intervals on eccentricity at $10$~Hz in the detector frame, $e_{10}$, against source-frame chirp mass $\mathcal{M}$ of the 62 BBH analysed in this work.
On the right-hand side, we plot median and upper $90\%$ credible intervals for $e_{10, {\rm peak, source}}$, the eccentricity at a peak gravitational-wave frequency of $10$~Hz in the source frame, following the conversions described above.
The number of binaries with $e_{10, {\rm peak, source}} \geq 0.05$ is the same as that with detector-frame eccentricity at $10$~Hz Keplerian frequency $e_{10} \geq 0.05$: we find four binaries above this threshold in both cases.

\begin{figure}
    \centering
    \includegraphics[width=0.45\textwidth]{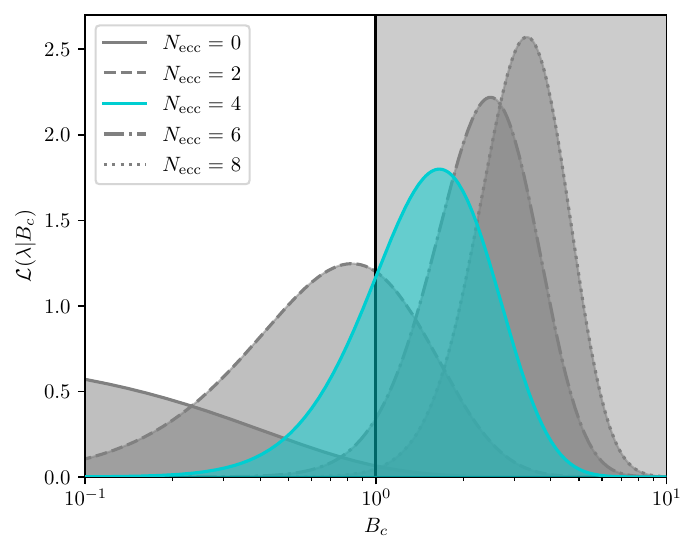}
    \caption{The posterior probability of the branching fraction, $B_c$, if all eccentric mergers are produced in globular clusters, following the model in \citet{Zevin:2021:seleccentricity}. We take $N_{\rm obs}=62$ as the number of observed events, not counting those that end up undersampled after reweighting. The black line denotes $B_c=1$, which corresponds to 100\% of mergers being formed in globular clusters. Above this line, non-zero likelihood support indicates that there are more eccentric mergers than can be accounted for with globular cluster mergers alone. We plot the probability for $N_{\rm ecc} = 4$, the number of mergers we find with $e_{10} \geq 0.05$, and for \new{$N_{\rm ecc} = 0, 2, 6$ and $8$, to demonstrate alternative scenarios that are possible given the caveats of our method. If we are mistaking higher-order modes or spin-precession for eccentricity, we may have fewer than four truly eccentric events, leading to a branching fraction likelihood that peaks below the upper prior limit of $B_c=1$. Alternatively, if the limitations of the reweighting method are causing eccentric events to remain undetected, we may have more than four truly eccentric events, leading to the majority of the branching fraction likelihood being outside of the range of the prior.} It remains feasible for $\leq100\%$ of observed mergers to be forming in globular clusters with four eccentric events within the population. However, the peak of the probability lies above the $B_c=1$ line, indicating that another process may be producing eccentric mergers. Additionally, if many of the candidates with less well-measured eccentricities are truly eccentric, the globular cluster hypothesis becomes less likely.}
    \label{fig:branching_fraction}
\end{figure}

\subsubsection{\newnew{Constraining the fractional contribution of mergers formed in dense star clusters}}

We can now use the fraction of binaries detected with $e_{10} \geq 0.05$ to constrain the fractional presence of mergers produced in dense star clusters within the population.
Consistent with \citet{Zevin:2021:seleccentricity}, we call this the branching fraction, $B_c.$
In Fig. \ref{fig:branching_fraction}, we plot in teal the likelihood on $B_c$ calculated following \citet{Zevin:2021:seleccentricity}, which treats the number of eccentric mergers in the observed population as a Poisson counting problem.
The likelihood for detecting four eccentric binaries within 62 mergers is represented by the teal curve, with the grey shaded rectangle indicating the region in which the prior probability goes to zero ($B_c > 1$).

Since we still have a small population and many uncertainties from both our method and our theoretical models, we cannot constrain $B_c$ very tightly or confidently at this stage.
\new{Since we require that the eccentric posterior probability distribution overlaps enough with the quasi-circular posterior probability distribution in order to retain a sufficient sample size after reweighting, we are biased towards detecting low- or negligible-eccentricity sources; high-eccentricity ($e_{10} > 0.2$) sources may be mistaken for quasi-circular mergers if they are not undersampled.
Therefore, we may be under-reporting the true number of eccentric mergers in the set of $62$ adequately-sampled events we consider.
Conversely, spin effects can mislead inferences of eccentricity \citep[e.g.,][]{JuanHeadOn, Romero-Shaw:2020:GW190521, OSheaKumar2021}; the signals that we infer to show the fingerprints of eccentricity may in fact be due to spin-induced precession.}
\textit{If} four of our eccentric candidates are truly eccentric, and \textit{if} all eccentric mergers are formed within globular clusters, then globular clusters must contribute at least $35\%$ of observed mergers, at $95\%$ credibility (to calculate this we use the posterior, i.e., we ignore parts of the likelihood greater than one).

\subsection{A population model for the eccentricity distribution}
\label{sec:population-model}

\begin{figure}
    \centering
    \includegraphics[width=0.45\textwidth]{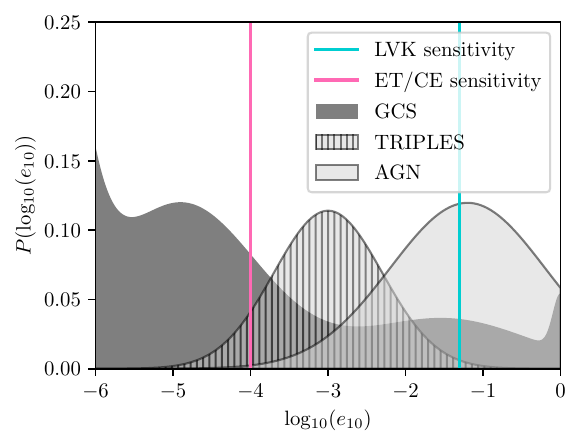}
    \caption{A simplistic toy model for the eccentricity distribution arising as a byproduct of three different formation mechanisms: dynamical mergers in GCs (dark grey) and AGN (light grey), in addition to field triples (striped grey). Each mechanism is represented by either a single Gaussian or a mixture of Gaussians of different weights. The vertical teal line shows the approximate limit of LIGO-Virgo sensitivity to eccentricity at $e_{10} = 0.05$ and the vertical pink line shows the estimated sensitivity of third-generation detectors Einstein Telescope (ET) and Cosmic Explorer (CE) to eccentricity at $e_{10} \approx 10^{-4}$. }
    \label{fig:toy_population}
\end{figure}

There are multiple pathways that can lead to BBH mergers with detectable eccentricity in the LIGO-Virgo sensitivity band.
As the population grows, it may be possible to distinguish mergers from these different channels by studying the shape of the population eccentricity distribution.
In Figure~\ref{fig:toy_population}, we illustrate a simplified population model containing three mechanisms that may produce eccentric mergers with distinct eccentricity distributions: mergers facilitated by interactions within globular cluster mergers (GCS), binary mergers occurring within field triples (TRIPLES), and mergers inside active galactic nuclei (AGN).
These populations are represented as follows:

\begin{itemize}
    \item \textbf{GCS}: We represent the distribution shown in Figure~1 of.   \cite{Zevin:2021:seleccentricity} using a Gaussian mixture model containing four Gaussians in log$_{10}(e_{10})$ at $\mu = [-6.9, -4.9, -1.5, 0]$ with widths $\sigma = [0.5, 1.0, 1.1, 0.1]$. These peaks, respectively, correspond to: mergers that form in clusters and are ejected before they merge; mergers that occur after dynamical interactions within clusters; gravitational-wave capture mergers; and gravitational-wave capture mergers that become bound within the LVK band, merging with eccentricities close to unity.
    \item \textbf{TRIPLES}: Binaries can be driven to merge rapidly in isolation if gravitational energy is removed from their orbit by a third object with which they are bound. In some cases, the eccentricity of these binaries can be amplified through Lidov-Kozai oscillations \citep[e.g.,][]{Lidov62, Kozai62}. Following \cite{Lower18} and references therein, we model the primary component expected eccentricity distribution from field triples as a single Gaussian in log$_{10}(e_{10})$ centred at $\mu = -3$ with width $\sigma = 0.7$ for simplicity. We ignore the small higher-eccentricity peak expected for $\sim 5\%$ of field triples since the contribution from this channel is expected to be small \citep[e.g.,][]{Silsbee16, RodriguezAntonini2018}, and we include it in our model for illustrative purposes only (we do not include it in our toy-population-model analyses).
    \item \textbf{AGN}: In the dense centre and accretion disk of an active galactic nucleus, binaries can be driven to merge through dynamical interactions. The eccentricity distribution of dynamical mergers inside AGN differs from that expected in globular clusters because the central gravitational potential of the nucleus is greater, and various interactions can also occur within the accretion disk. The distribution expected from AGN is highly uncertain and depends greatly on the properties of the AGN. With reference to \cite{Samsing:2020:AGNeccentric} (see the right-hand panel of their Figure 5 in particular), we represent the contribution from AGN as a Gaussian in log$_{10}(e_{10})$ centred at $\mu = -1.2$ and $\sigma = 1.0$. We recognise this as a greatly simplified model, and present this as a proof-of-concept that can be extended to include more complete distributions predicted by complex AGN models.
\end{itemize}

\begin{figure}
    \centering
    \includegraphics[width=0.5\textwidth]{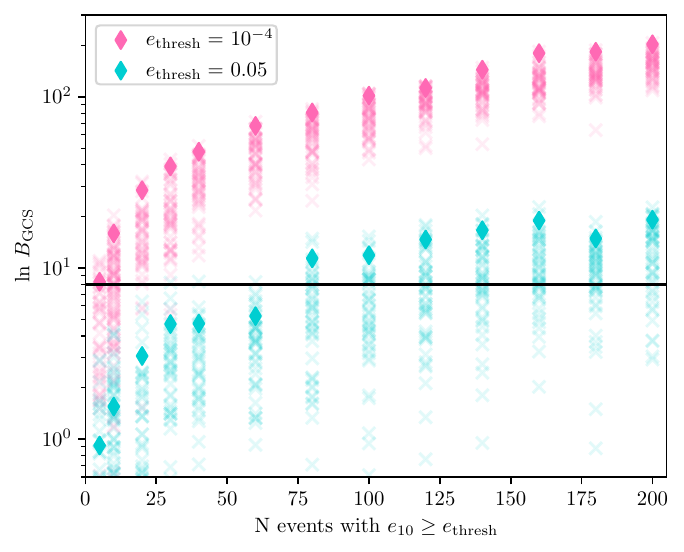}
    \caption{The ln Bayes factor ln~$B_{\rm GCs}$ for the GC hypothesis relative to the AGN hypothesis with $N$ events and studying eccentricities only above $e_{\rm thresh}$. The two $e_{\rm thresh}$ represent the approximate sensitivity of current detectors ($e_{10} \sim 0.05$) and third-generation detectors ($e_{10} \approx 10^{-4}$). 
    For each $N$, we calculate ln~$B_{\rm GCs}$ for 50 mock populations (shown with translucent crosses) and take the average (shown with opaque diamonds). The threshold value of ln~$B_{\rm GCs}$ for ``confident'' detection, which we take to be $8$ by convention, is represented with a black line.}
    \label{fig:log_B_GCs}
\end{figure}
Inferring the formation mechanisms of binaries should incorporate contributions from multiple different channels, likely many more than are represented in our simple model. 
For the sake of this proof-of-principle demonstration, we generate populations of events from the GCs channel only, approximating each event's eccentricity posterior probability distribution as a delta function. 
Since our eccentricity measurements typically have broader posteriors than a delta function, the result of this simple demonstration should be taken as a lower limit.
We generate populations ranging in size from $5$ to $200$ events, and recover the population distribution using the \texttt{hyper} submodule of \texttt{bilby} \citep{bilby, Romero-Shaw:2020:Bilby}.
We obtain ln Bayes factors, ln~$B_{\rm GCS}$, for the strength of the GC hypothesis relative to the AGN hypothesis, taking ln~$B_{\rm GSs}\geq8$ as a ``confident'' preference for the GC hypothesis.
We study only the range of eccentricities that is accessible to each era of detector:

\begin{itemize}
    \item \textbf{LVK sensitivity limit}: Although sensitivity to eccentricity varies slightly with mass (see Appendix C of \cite{Romero-Shaw:2021:GWTC-2-ecc}), we take $e_{10} \geq 0.05$ as the range within which existing detectors can measure eccentricity. This is consistent with our own measurements, as well as the predictions of \cite{Lower18}.
    \item \textbf{Einstein Telescope / Cosmic Explorer sensitivity limit}: We use samples from the range of eccentricities to which the Einstein Telescope (ET) and Cosmic Explorer (CE) are expected to be sensitivity; this is approximately $e_{10} \geq 10^{-4}$ \citep{Lower18}.  
\end{itemize}

For each $N$ and for each sensitivity threshold, we perform parameter estimation on $50$ mock populations. 
We plot our results in Figure~\ref{fig:log_B_GCs}, where the average ln~$B_{\rm GCS}$ for each $N$ obtained is marked with a diamond.
We find that we will be able to confidently distinguish a BBH merger population dominated by those formed in GCs from one in which mergers form in AGN with $\gtrsim 80$ detectably-eccentric events.
This is a lower limit, since real detections have non-negligible uncertainties.
Additionally, it will be trickier to disentangle the contributions of multiple channels that contribute comparable numbers of eccentric binaries to the population.
We leave a detailed simulation study that accounts for these complications for future work.

\section{Conclusion}
\label{sec:conclusion}

In this work, we analyse $26$ BBH signal candidates \new{newly added to the LVK catalogue of gravitational-wave transients in GWTC-2.1 and GWTC-3} for signs of orbital eccentricity using an aligned-spin, eccentric waveform model. We find that two of these events have significant support for detectable eccentricity at $10$~Hz: GW191109 and GW200208\_22. Together with two events from GWTC-2, GW190521 and GW190620 \citep{Romero-Shaw:2021:GWTC-2-ecc}, four of the \Nevents~BBH we have studied show significant support for measurable eccentricity. At just over $6\%$, this is slightly more than the $\sim 4\%$ of mergers that should be expected with detectable eccentricity from GCs \citep{Zevin:2021:seleccentricity}, and may indicate that there are more eccentric mergers in the population than can be explained by dense star clusters alone. However, as we show in Section \ref{sec:population}, the difference can be explained by Poisson noise. Additionally, some of these mergers may be spin-precessing binaries that are masquerading as eccentric BBH in our analyses due to our enforced assumption of spin-alignment. 

While we cannot say conclusively that these four binaries are eccentric and spin-aligned, our results do show that these signals are consistent with those that contain hints of orbital eccentricity, and are better-fit by eccentric waveforms than they are by quasi-circular waveforms. If they \textit{are} eccentric, and/or have large spin magnitudes and/or spin tilts, then these binaries are difficult to explain through isolated evolution scenarios and add weight to the hypothesis that the BBH detected by the LVK are dynamically formed. 

We have focused on the particular dynamical formation environment of globular star clusters in this paper.
However, we emphasise that this is due to the robustness of predictions from this channel, as opposed to any sign that this particular dynamical formation environment is preferred over other dynamical formation environments. 
In fact, this formation environment may be particularly \textit{unlikely} when other parameters of these systems are considered.
Three of the four potentially eccentric events have notably high primary masses, all with median source-frame measurements above the tentative pair-instability supernova mass gap lower limit of $\sim55$ M$\odot$ \citep[e.g.,][]{HegerWoosley02, Belczynski16, Woosley17, Marchant:2020:MassGap} (the exception, GW200208\_22, has a median source-frame primary mass of $\sim 25$ M$_\odot$ after reweighting to the eccentric model). 
The relatively low escape velocities of globular clusters \citep[typically $\lesssim 120$~km~s$^{-1}$;][]{Gnedin:2002:MWGCVesc, AntoniniRasio:2016:GCNSCVesc, Baumgardt:2018:MWGCVesc} mean that merger remnants are more often kicked out of the cluster than in environments with deeper central potential wells, such as AGN or nuclear star clusters \citep[e.g.,][]{AntoniniRasio:2016:GCNSCVesc, Fragione:2020:GW190521StarClusersHierarcical, Ford:2021:LoudVsQuiet, Fragione:2022:NSCs, Parthapratim:2021:HierarchicalKicks}.
Additionally, because globular clusters undergo mass segregation soon after their formation at high redshift, it is likely that the heaviest binaries merge outside of the observable range of current detectors \citep{AntoniniRasio:2016:GCNSCVesc, Romero-Shaw:2021:GCs}.
To assess the probability of formation in globular clusters relative to the probability of formation in AGN, we would need a multidimensional version of the toy-model analysis demonstrated in Section \ref{sec:population-model} incorporating robust predictions for eccentricity, mass, spin and redshift from these different environments; see , e.g., \citet{Yang:2019:HierarchicalAGN, Samsing:2020:AGNeccentric, McKernan:2020:AGNPredictions, Tagawa:2020:AGNdiskspin, Tagawa:2021:massgap, Tagawa:2021:AGN, Vajpeyi:2021:AGN, Gayathri:2021:AGN} for a range of recent predictions for binary black hole mergers occurring in AGN.

There are many roads to forming merging BBH dynamically, and we show in this paper that in a future when eccentricity can be tightly constrained, it will be possible to disentangle contributing dynamical channels with $\gtrsim 80$ detectably-eccentric mergers at current detector sensitivity limits. As detectors improve, the threshold for detectable eccentricity decreases, so fewer detectably-eccentric events are required to distinguish contributions from different dynamical channels. Even so, without a method to confidently measure eccentricity and the full range of spin effects simultaneously, it will not be possible to use the method we have demonstrated to identify the dominant dynamical formation channel. We will therefore turn our focus towards extricating measurements of eccentricity from the restraints of spin-aligned, moderately-spinning waveform models in the future.

\section{Acknowledgements}
We thank Vijay Varma, Katerina Chatziioannou, Alan Knee, Teagan Clarke, Makai Baker, and attendees of the Niels Bohr Institute's Workshop on Black Hole Dynamics for enlightening suggestions, observations, and discussions.
We also thank Mike Zevin for his insights on our work and comments on our manuscript.
\new{We are grateful to our anonymous reviewer, whose advice on our manuscript improved its contents.}
This work is supported through Australian Research Council
(ARC) Centre of Excellence CE170100004, and Discovery Project
DP220101610.
IMR-S acknowledges support received from the Herchel Smith Postdoctoral Fellowship Fund.
The authors are grateful for computational resources provided by the LIGO Laboratory and supported by National Science Foundation Grants PHY-0757058 and PHY-0823459.
This research has made use of data or software obtained from the Gravitational Wave Open Science Center (gw-openscience.org), a service of LIGO Laboratory, the LIGO Scientific Collaboration, the Virgo Collaboration, and KAGRA. LIGO Laboratory and Advanced LIGO are funded by the United States National Science Foundation (NSF) as well as the Science and Technology Facilities Council (STFC) of the United Kingdom, the Max-Planck-Society (MPS), and the State of Niedersachsen/Germany for support of the construction of Advanced LIGO and construction and operation of the GEO600 detector. Additional support for Advanced LIGO was provided by the Australian Research Council. Virgo is funded, through the European Gravitational Observatory (EGO), by the French Centre National de Recherche Scientifique (CNRS), the Italian Istituto Nazionale di Fisica Nucleare (INFN) and the Dutch Nikhef, with contributions by institutions from Belgium, Germany, Greece, Hungary, Ireland, Japan, Monaco, Poland, Portugal, Spain. The construction and operation of KAGRA are funded by Ministry of Education, Culture, Sports, Science and Technology (MEXT), and Japan Society for the Promotion of Science (JSPS), National Research Foundation (NRF) and Ministry of Science and ICT (MSIT) in Korea, Academia Sinica (AS) and the Ministry of Science and Technology (MoST) in Taiwan.

\bibliography{bib}

\appendix

\section{Likely non-eccentric \new{new events in GWTC-3}}
\label{sec:noneccentric}

\begin{table*}
\centering
\caption{A summary of the detector-frame eccentricity measurements for \new{newly-studied} events in GWTC-3 with negligible support for the hypothesis that $e_{10} \geq 0.05$.
Columns are as described in the caption for Table \ref{tab:results-eccentric}. 
\label{tab:results-not-eccentric}}
\bgroup
\def\arraystretch{1.5}
\begin{tabular}{c|c|c|c|c|c}
Event name & $e_{10} \geq 0.1$ (\%) & $e_{10} \geq 0.05$ (\%) & $\ln \mathcal{B} (e_{10} \geq 0.1)$  & $\ln \mathcal{B} (e_{10} \geq 0.05)$ & $n_{\rm eff}$ \\
\hline
GW190426\_16 & $3.25$ & $10.69$ & $-0.86$ & $-0.54$ & $1170$ \\
GW190916 & $4.55$ & $11.71$ & $-0.59$ & $-0.35$ & $573$ \\
GW190926 & $6.75$ & $15.61$ & $-0.15$ & $-0.10$ & $42171$ \\
GW191103 & $9.20$ & $17.82$ & $-0.14$ & $-0.09$ & $175$ \\
GW191222 & $0.00$ & $9.40$ & $-1.68$ & $-0.96$ & $118$ \\
GW191230 & $6.75$ & $14.35$ & $-0.46$ & $-0.34$ & $10661$ \\
GW200112 & $0.77$ & $7.94$ & $-0.88$ & $-0.50$ & $39271$ \\
GW200128 & $2.66$ & $11.60$ & $-0.83$ & $-0.36$ & $11755$ \\
GW200208 & $3.57$ & $11.86$ & $-0.70$ & $-0.42$ & $53832$ \\
GW200219 & $3.50$ & $11.53$ & $-0.74$ & $-0.43$ & $36630$ \\
GW200220\_O6 & $4.19$ & $13.16$ & $-0.57$ & $-0.31$ & $20910$ \\
GW200220\_12 & $7.47$ & $16.20$ & $-0.26$ & $-0.14$ & $13586$ \\
GW200224 & $4.34$ & $16.17$ & $-0.52$ & $-0.03$ & $38653$ \\
GW200302 & $2.44$ & $9.15$ & $-1.08$ & $-0.38$ & $164$ \\
GW200306 & $5.73$ & $14.18$ & $-0.44$ & $-0.25$ & $4995$ \\
GW200308 & $7.60$ & $16.24$ & $-0.03$ & $-0.03$ & $672$ \\
GW200311 & $0.78$ & $6.70$ & $-1.61$ & $-0.83$ & $51442$ \\
GW200316 & $0.77$ & $18.46$ & $-0.67$ & $-0.02$ & $131$ \\
\end{tabular}
\egroup
\end{table*}

\begin{figure*}
    \centering
   \includegraphics[width=0.6\textwidth]{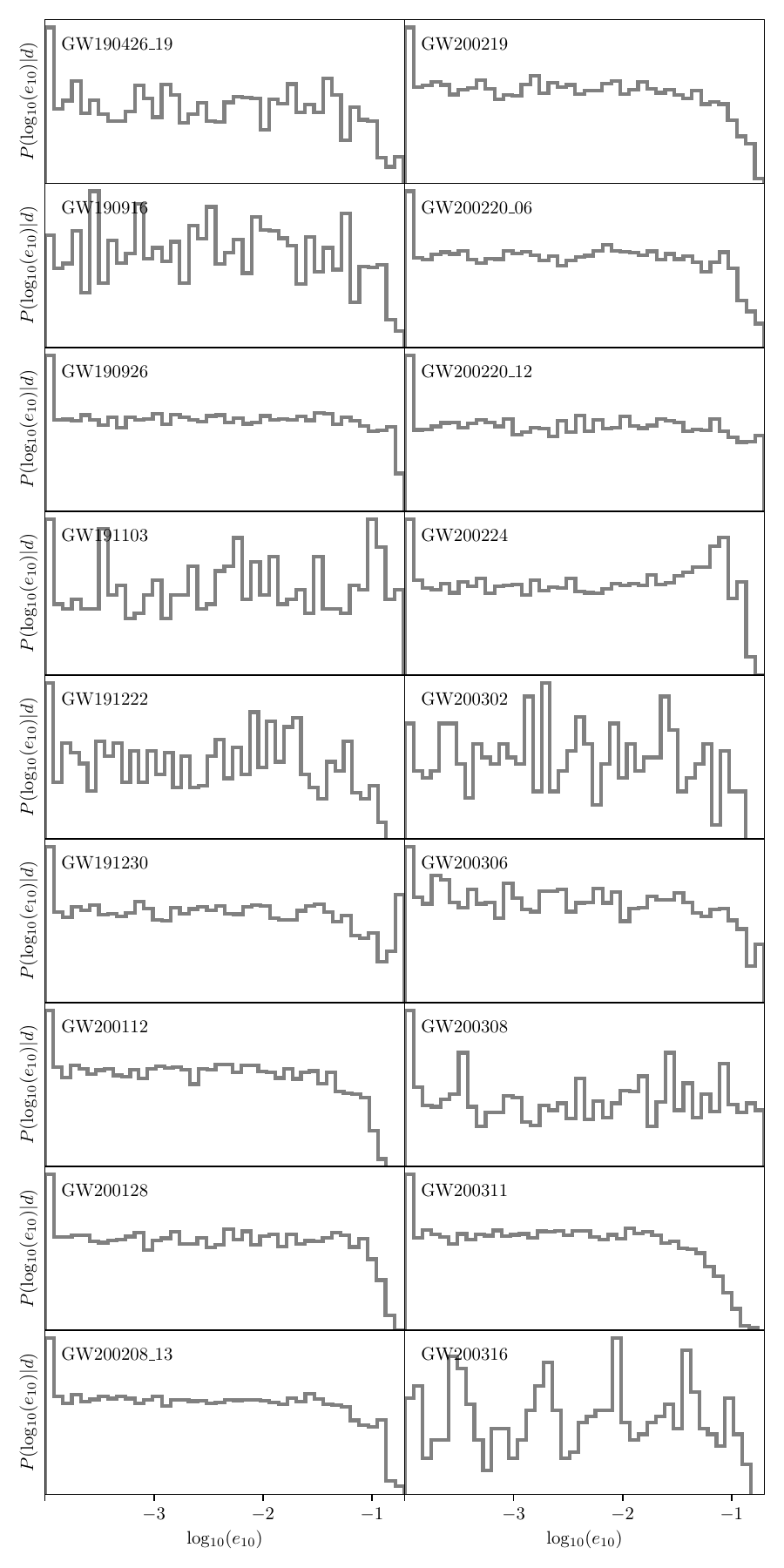}
    \caption{Marginal posterior distributions on log$_{10}(e_{10})$ for \new{newly-studied} events in GWTC-3 that have negligible support for $e_{10} \geq 0.05$. We label each panel with the name of the event.}
    \label{fig:not_eccentric_posteriors}
\end{figure*}

In Table \ref{tab:results-not-eccentric}, we provide summary statistics for the \new{newly-studied events in GWTC-3} that are both adequately sampled and have $\ln \mathcal{B} (e_{10} \geq 0.05) < 0$.
Marginal posterior distributions on log$_{10}(e_{10})$ are displayed in Figure \ref{fig:not_eccentric_posteriors}.
Full posterior distributions on all parameters of these events are provided online.\footnote{\href{https://github.com/IsobelMarguarethe/eccentric-GWTC-3}{github.com/IsobelMarguarethe/eccentric-GWTC-3}}

\section{Undersampled events in GWTC-3}
\label{sec:undersampled}

\begin{figure}
    \centering
    \includegraphics[width=0.45\textwidth]{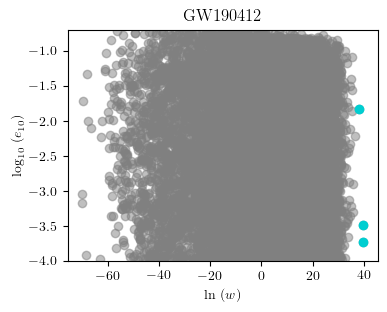}
    \includegraphics[width=0.45\textwidth]{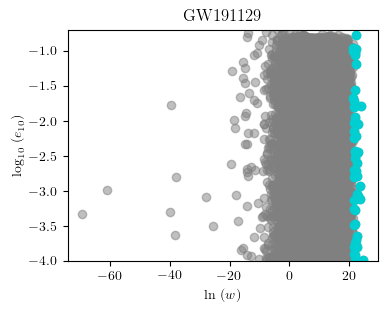}
    \includegraphics[width=0.45\textwidth]{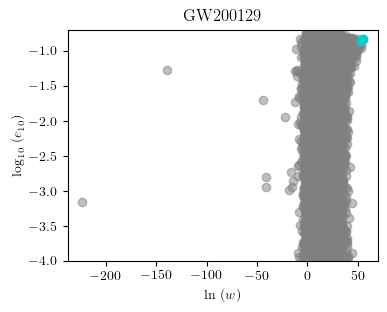}
    \caption{Scatter plots of natural log weight against log$_{10}~(e_{10})$ for three example undersampled events. We label each panel with the name of the event. In each plot, the $n_{\rm eff}$ most highly-weighted samples are highlighted as teal dots. These dots are the points that dominate the reweighted posterior. By plotting the weights against eccentricity in this way, we can see trends that hint at the reason for undersampling. GW190412 \new{(3 effective samples; top left panel)} and GW191129 \new{(49 effective samples; top right panel)} have highly-weighted samples spread across the eccentricity range, and have relatively low weights for highly-eccentric samples. This hints at waveform systematics being the root of undersampling; if we removed the $n_{\rm eff}$ most highly-weighted samples from these distributions, the resulting posterior on eccentricity would be mostly uninformative, with decreasing support above $e_{10} \approx 0.1$. Meanwhile, the plot for GW200129 (\new{1 effective sample; bottom panel}) shows an overall trend towards higher weights for higher eccentricities. If we removed the most highly-weighted sample from this distribution, the posterior would still be undersampled, with the most highly-weighted sample remaining above $e_{10} = 0.1$. To obtain a well-sampled posterior, we would need to remove most of the samples at high eccentricities, hinting that there is strong support for eccentricity in the data and that the eccentric posterior does not have enough overlap with the quasi-circular posterior for the former to be well-sampled using the reweighting method.}
    \label{fig:weights_scatter_plots}
\end{figure}

\new{We consider any event with $n_{\rm eff} < 100$ to be undersampled. With such few samples, measurements can be misleading, so we do not include in this work the analyses of the events in GWTC-3 for which the reweighting process is unsuccessful.} \new{However, even when an event is undersampled}, it can be informative to study a scatter plot of weights versus eccentricities to gain an understanding of the reason for undersampling. 
Three examples are shown in Figure \ref{fig:weights_scatter_plots}, in which the $n_{\rm eff}$ (see Eq. \ref{eq:neff}) dominating samples are highlighted. 
Firstly, the locations of the scatter points across the eccentricity range are informative: if the event has no support at all for high eccentricities, it will not be possible for highly-eccentric samples to be drawn, so there will be no scatter points at high eccentricities.

Secondly, the distribution of weights across the eccentricity range is informative.
If highly-weighted samples are spread evenly across the range of eccentricities---as is the case for our first two examples---waveform systematics are a probable suspect for underweighting: these samples are likely to reside in an area of the wider parameter space (for example, a particular combination of masses and spins) for which the \texttt{IMRPhenomD} waveform does not well-represent the $e_{10} = 0$ \texttt{SEOBNRE} waveform.
If instead the dominating samples are localised to a particular part of the eccentricity range---as is the case for our final example, GW200129---the situation is both more interesting and more frustrating.
In this case, it is likely that there is strong support for eccentricity in the data, but that the eccentric posterior does not overlap the quasi-circular posterior to an adequate extent for the reweighting method to be efficient.
\end{document}